\documentclass[aps,pra,reprint,showpacs,superscriptaddress]{revtex4-1}
\usepackage{amsthm}
\usepackage{epsfig}
\usepackage[]{natbib}
\usepackage{xspace}
\usepackage{color}
\usepackage{bbold}
\usepackage{bm}
\usepackage{bbm}
\usepackage{tabu}
\usepackage{physics}
\usepackage{mathtools}
\usepackage[colorlinks=true,
			urlcolor=blue,
            linkcolor=blue,
            anchorcolor=blue,
            citecolor=blue
            ]{hyperref}

\newtheorem*{remark}{Remark}

\begin{document}

\title{
Large fluctuations of the first detected quantum return time
 }

\author{R. Yin} 
\affiliation{Department of Physics, Institute of Nanotechnology and Advanced Materials, Bar-Ilan University, Ramat-Gan 52900, Israel}
\author{K. Ziegler}
\affiliation{Institut f\"ur Physik, Universit\"at Augsburg, $D-86135$ Augsburg, Germany}
\author{F. Thiel}
\affiliation{Department of Physics, Institute of Nanotechnology and Advanced Materials, Bar-Ilan University, Ramat-Gan 52900, Israel}
\author{E. Barkai}
\affiliation{Department of Physics, Institute of Nanotechnology and Advanced Materials, Bar-Ilan University, Ramat-Gan 52900, Israel}

\begin{abstract}
How long does it take a quantum particle to return to its origin?
As shown previously under repeated projective measurements aimed to detect the return,
the closed cycle yields a geometrical phase which 
shows that the average first detected return time is quantized.
For critical sampling times or when parameters of the Hamiltonian are tuned 
this winding number is modified.
These discontinuous transitions exhibit gigantic fluctuations of the return time.  
While the general formalism of this problem was studied at length, 
the magnitude of the fluctuations, 
which is quantitatively essential, 
remains poorly characterized. 
Here, we derive explicit expressions for the variance of the return time, 
for quantum walks in finite Hilbert space.
A classification scheme of the diverging variance is presented, 
for four different physical effects: 
the Zeno regime, 
when the overlap of an energy eigenstate and the detected state is small  
and when two or three phases of the problem merge. 
These scenarios present distinct physical effects 
which can be analyzed with the fluctuations of return times investigated here, 
leading to a topology-dependent time-energy uncertainty principle.
\end{abstract}
\maketitle

\section{Introduction}
The return time is the time it takes a quantum \cite{Gruenbaum2013,Bourgain2014} or classical particle \cite{Redner2001,benichou2015} to return to its origin. For diffusive particles this time determines chemical reaction
rates while for celestial mechanics the return of a comet is a classical problem.
For a quantum particle, say on a graph, the definition of the return
time needs the introduction of a measurement protocol.
A well investigated approach is to consider measurements on the original node, repeated
stroboscopically until the first detection. 
The problem is to find the distribution of the number of attempts $n$ till first detection \cite{Krovi2006,Krovi2006a,Krovi2007,Krovi2008,Stefanak2008,Krapivsky2014,Dhar2015,Sinkovicz2016,Friedman2017a,Friedman2017,Thiel2018a,Lahiri2019,Ambainis2001,Nitsche}.
This sheds light on the backfire of measurement on unitary evolution, 
and on time processes in quantum mechanics. 
Besides its fundamental aspect this line of research became important in the context of quantum search \cite{Grover1997,Childs2004,LiShanshan}, where one of the basic questions is whether repeated measurements destroy or enhance success of quantum search \cite{Krovi2006,Krovi2006a,Krovi2007,Krovi2008,Ambainis2001,BACH2004}.
For that one would like to know the average of $n$ and its fluctuations.

Recently, Gr\"{u}nbaum {\it et al} \cite{Gruenbaum2013,Bourgain2014} demonstrated 
how this problem is related to a new class of geometrical phases. 
In the quantum return problem the particle performs a loop, as it starts 
and is measured on the same spot, so a cycle is found. 
However, unlike other approaches \cite{ABeffect,Berry1984,QNiu2010,Flurin2017}
to topology here unitary evolution is pierced by measurements.
The result is that the average of $n$ is equal to a winding number $w$, which in turn is equal to the number of distinct phases $\exp(iE_k\tau)$ in the system. 
Here $E_k$ are the energy levels of the time-independent Hamiltonian $H$, and $\tau$ is the sampling time, $\hbar$ is set to $1$ in this manuscript.
So $\langle n \rangle = w$ is quantized.
%
\begin{figure}[htbp]
\centering
\includegraphics[width=1.0\columnwidth]{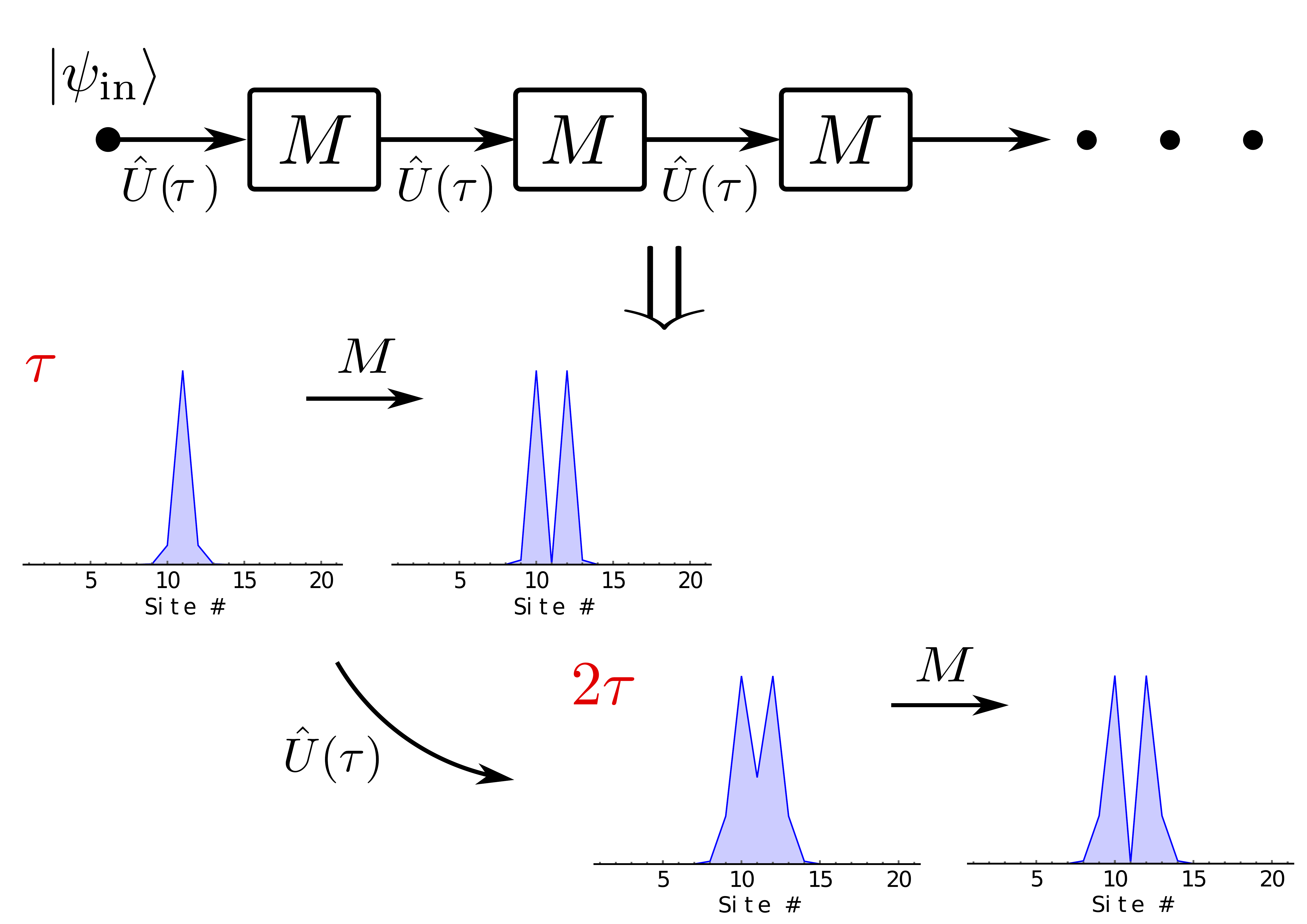}
\caption{(Color online)
A sketch of the measurement protocol. Measurements $M$ are performed every $\tau$ units of time, between which the system evolves unitarily with $\hat{U}(\tau)$. 
In the lower part of the figure we present the effect of collapse, with failed detection. 
The system in mind is a quantum walk on a line with $21$ sites 
and the measurements are made on site $x=11$.  
The process is continuous until the first successful measurement.
From collapse theory we get a wipe-out of the wave function on this site
when the particle is not detected. 
This is shown in the sub-figures pointed by the arrows with a label $M$.
}
\label{fig0}
\end{figure}

The goal of this paper is to investigate the fluctuations
of the number of measurements needed till the first return is recorded, namely the variance of $n$. 
It was shown previously that these can be very large close
to critical sampling parameters even for small systems \cite{Gruenbaum2013,Friedman2017a}.
Here we provide formulas for the blow-ups of the fluctuations,
using  an elegant mapping of the quantum problem to a classical
charge theory \cite{Gruenbaum2013}. 
For critical values of the Hamiltonian's parameters or the sampling time, 
one finds discontinuities in the mean detected return time, 
namely $w \to w + {\cal \alpha}$ where $\alpha$ is a non-zero integer,
typically in the examples studied below $\alpha = \pm 1 \text{ or } \pm 2$.
Close to these non-analytical points, 
the fluctuations diverge.
We classify different scenarios for critical behavior, 
using a one, two, three or many-charge theory,
the physics in each case yields very different insights on the mechanism of the fluctuations' blow-ups.
As we discuss below, from an engineering point-of-view, the critical fluctuations are a nuisance since they deem the quantum search as non-practical,  
and our work shows how to avoid them.

\section{Model and Formalism}

\subsection{Model}
The system is initially prepared
in state $\ket{\psi_{{\rm in}}}$, for example on a node of a graph. 
Every $\tau$ units of time we perform a measurement
in an attempt to detect the particle in its initial state. 
Between the　measurements the evolution is unitary with $\hat{U}(\tau)= \exp( - i H \tau)$.
The Hamiltonian $H$ is time-independent, 
and we assume a finite Hilbert space, 
so energy levels are discrete. 
While the main results presented below are general as examples we will consider Hamiltonians
describing a particle hopping on a ring and a model system describing two interacting bosons 
in the process of tunneling.  
The measurements are strong, 
performed via the projector $\hat{D}=|\psi_\text{in}\rangle \langle \psi_\text{in}|$,
namely we use the collapse postulate.
The probability of detecting the particle at some time $t$ is as usual 
the squared absolute value of the amplitude of finding the particle in the detected state. 
For example if we measure on a node $\ket{r}$ of a graph, 
this is given by $|\langle r|\psi\rangle|^2$ 
and since we are dealing here with the return problem we also have 
$|\psi_{\rm in}\rangle = \ket{r}$.  
If the particle is detected, we are done. 
If not, the amplitude on the detected state is zero and the wave function is renormalized 
(see details in \cite{Friedman2017a} and schematics of the process in Fig. \ref{fig0}).  
The outcome of this procedure is a string of measurements ``failure, failure, $\cdots$'' and in the $n$-th attempt a ``success''. The time $n \tau$ is called the first detected 
return time which is random. 
~\\

\subsection{Recap: general formalism}
After presenting the model, 
we provide a primer on the quantum return time 
re-deriving the results obtained in \cite{Gruenbaum2013} 
using standard mathematical tools.
The first detection amplitude yields the statistics of the problem and is denoted $\phi_n$ \cite{Gruenbaum2013,Dhar2015,Friedman2017a}. We now discuss its properties.
The probability to detect the particle for the first time at the $n$-th attempt is the squared absolute value of this amplitude: $F_n = |\phi_n|^2$,
and the mean number of detection attempts till success is $\expval{n} = \sum n F_n$.
We remark that the normalization condition $\sum_{n=1}^\infty F_n =1$, 
is valid for finite Hilbert space and for the return problem under investigation \cite{Gruenbaum2013,thiel2019a}.
The amplitude $\phi_n$ is given by \cite{Gruenbaum2013,Dhar2015,Friedman2017a} 
\begin{equation}
	\phi_n = 
			\bra{ \psi_{{\rm in}} } 
			\big[
				\hat{U}(\tau)\hat{P}
			\big]^{n-1}
			\hat{U}(\tau)
			\ket{\psi_{{\rm in}}}, \quad \text{ $\hat{P}=\mathbb{1}-\hat{D}$.}
\label{eqSM01a}
\end{equation}
%
It shows that the unitary evolution, represented by $\hat{U}(\tau)$, 
is interrupted by projective measurements 
via the operation $(\mathbb{1}-\hat{D})$ for $(n-1)$ times 
until the $n$-th success.

A useful tool is the discrete Fourier transformation of $\phi_n$, 
%
\begin{equation}\label{FourierPhi}
    	\tilde{\phi}(e^{i\omega}) 
    							:=
    							\sum_{n=1}^{\infty}
    							e^{i \omega n} \phi_{n}
    							=
    							\langle \psi_{\rm in}| (e^{i \tau H - i \omega}-\hat{P})^{-1}| \psi_{\rm in}\rangle.
\end{equation}
%
Using the identity $(1+B)^{-1}=1-B(1+B)^{-1}$ ($B$ is a matrix) we get
\begin{equation}
\begin{aligned}
    \tilde{\phi}(e^{i\omega})
    						=& \langle \psi_{\rm in}| (e^{i \tau H - i \omega}-\hat{P})^{-1} | \psi_{\rm in}\rangle \\
    						=& \langle \psi_{\rm in}| (e^{i \tau H-i \omega} - \mathbb{1})^{-1} | \psi_{\rm in}\rangle \\
    						&- \langle \psi_{\rm in}| (e^{i \tau H-i \omega} - \mathbb{1})^{-1}| \psi_{\rm in}\rangle \tilde{\phi}(e^{i\omega})，
\end{aligned}
\end{equation}
or equivalently
\begin{equation}
    \tilde{\phi}(e^{i\omega})
    						= \frac{\langle \psi_{\rm in}| (e^{i \tau H-i \omega}-\mathbb{1})^{-1}
    						  |\psi_{\rm in}\rangle} 
    						  {1+\langle \psi_{\rm in}| (e^{i \tau H-i \omega}-\mathbb{1})^{-1}
    						  |\psi_{\rm in}\rangle}.
\end{equation}
Now we notice that 
$\langle \psi_{\rm in}| (e^{i \tau H-i \omega}-\mathbb{1})^{-1}|\psi_{\rm in}\rangle = \sum_{n=1}^\infty u_n e^{in\omega}$ where $u_n$ is defined:
\begin{equation}
    u_n := \langle \psi_{\rm in}| e^{-i n H \tau}| \psi_{\rm in}\rangle
    	=\sum_{k=1}^{w} p_k e^{-inE_k\tau}.
    \label{un}
\end{equation}
$u_n$ is the return amplitude describing dynamics free of measurements, 
i.e. $\exp( -i n H \tau) \ket{\psi_\text{in}}$ is the system's wave function at time $n \tau$ in the absence of measurements. 
Here we define the overlaps 
$p_k = \sum^{g_k}_{l=1}|\bra{\psi_{{\rm in}}} E_{kl} \rangle|^2 $, 
and $\{ \ket{E_{kl}} \}$ are the eigenstates of
$H$ corresponding to the eigenvalue $E_k$ with $0<l\le g_k$, 
where $g_k$ is the degeneracy. 
$w$ is the number of distinct phases $e^{iE_k\tau}$ with corresponding non-zero $p_k$.
This means that if we modify $\tau$ in such a way that 
$\exp(i E_k \tau) = \exp(i E_{k'} \tau)$ for $k \neq k'$ $w$ will be reduced by one.
This effect is called merging of phases \cite{Gruenbaum2013}. 
In a disordered system with no degeneracy in the sense  
$\exp(i E_k \tau) \neq \exp( i E_{k'} \tau)$ for all energy states, 
and when all $p_k$ are non-zero, 
$w$ is the dimension of the Hilbert space. 
In general 
\begin{equation}\label{eq6}
\begin{aligned}
	\tilde{u}(e^{i\omega})	:=& \sum_{n=1}^\infty e^{in\omega} u_n 
							= \langle \psi_{\rm in}| (e^{i \tau H-i \omega}-\mathbb{1})^{-1}|\psi_{\rm in}\rangle \\
							=& \sum_{k=1}^w {p_k \over e^{iE_k\tau-i\omega}-1}.
\end{aligned}
\end{equation}
With this formula, $\tilde{\phi}(e^{i\omega})$ can be re-expressed as
\begin{equation}
\begin{aligned}
\tilde{\phi}(e^{i\omega}) &=	{\tilde{u}(e^{i\omega}) \over 1+\tilde{u}(e^{i\omega})}
						&=	{   
							e^{i\omega} \sum_{k=1}^w { p_k / (e^{i E_k \tau} - e^{i\omega})} 
							\over
							\sum_{k=1}^w { p_k  e^{i E_k \tau} / (e^{i E_k \tau} - e^{i\omega})}
							}.
\end{aligned}
\label{eqSM02}
\end{equation}
One can easily find that $[\tilde{u}(e^{i\omega})]^\ast=-[1+\tilde{u}(e^{i\omega})]$ or $|\tilde{\phi}(e^{i\omega})| = 1$, and this in turn gives the normalization $\sum_n F_n =1$.

We now switch $\exp(i \omega) \to z$ 
and write $\tilde{\phi}(z)= {\cal N}(z)/{\cal D}(z)$. 
$\tilde{\phi}(z)$ is called the generating function of $\phi_n$ \cite{Gruenbaum2013,Friedman2017a}.
The numerator ${\cal N}(z)$ and the denominator ${\cal D}(z)$ are related. 
Using Eq. (\ref{eqSM02}), it is easy to show that 
%
\begin{equation}\label{ND}
\begin{aligned}
\mathcal{N}(z) :=& \tilde{u}(z)  \prod_k^w(e^{iE_k\tau}-z) 
			    = 	
							z  \sum_{k=1}^{w} p_k
							\prod_{\substack{j=1 \\ j \neq k}}^{w}(e^{iE_j  \tau}-z) 
					, \\
\mathcal{D}(z) :=& [1 + \tilde{u}(z)]  \prod_k^w(e^{iE_k\tau}-z) 
				= 	
							\sum_{k=1}^{w} p_k \, e^{iE_k  \tau}
							\prod_{\substack{j=1 \\ j\neq k}}^{w}(e^{iE_j  \tau}-z) 
					.
\end{aligned}
\end{equation}
A straightforward calculation shows 
\begin{equation}
{\cal D}(z) = \exp(i \gamma) (-1)^{w-1} z^w \left[ {\cal N} \left( {1 \over z^*}\right) \right]^*,
\label{eqS04}
\end{equation}
where $\gamma=\sum_{k=1}^w E_k \tau$.
Clearly ${\cal N}(z)$ is a polynomial which is now rewritten as
\begin{equation}
{\cal  N} (z) = z  \prod_{i =1} ^{w-1} (z_i - z),
\label{eqSM05a}
\end{equation}
where $\{ z_i \}$ are the zeros of $\tilde{\phi}(z)$,
which are complex numbers within the unit disk.
Mathematically these zeros determine the fluctuations of $n$, see Eq. (\ref{eq02}) below.
Using Eq. (\ref{eqS04}) we find a useful identity \cite{Gruenbaum2013,Friedman2017a}: 
%
\begin{equation}
\tilde{\phi}(z) = 
				e^{- i \gamma} 
				{ 
					z \prod_{i=1} ^{w-1} (z_i -z) 
					\over
					\prod_{i=1} ^{w-1} (1 - z_{i}^* z )
				}.
\label{eqSM05}
\end{equation}
%
We will discuss the zeros $\{ z_{i} \}$ more carefully soon, 
as they are key to our main results. 

In this paper we investigate the fluctuations of $n$ and for that aim we now find the moment-generating function which is the discrete Fourier transform of $F_n$. 
Applying the Fourier transform to the first detected return probability 
and using $F_{n}=|\phi_n|^2$:
\begin{widetext}
\begin{eqnarray}\label{fr}
	\tilde{F}(\varphi) 
		&:=& \sum_{n=1}^\infty e^{in\varphi} |\phi_{n}|^2 = \sum_{n=1,m=1}^\infty e^{in\varphi} \phi_{n}\phi^\ast_{m}\delta_{nm} \nonumber 
		= {1 \over 2\pi} \int_{-\pi}^{\pi} \sum_{n=1}\phi_{n}e^{in(\omega+\varphi)}\, \sum_{m=1}\phi^\ast_{m} e^{-im\omega} \, d\omega  \nonumber \\
		&=& {1 \over 2\pi} \int_{-\pi}^{\pi}\tilde{\phi}(e^{i(\omega+\varphi)})\, [\tilde{\phi}(e^{i\omega})]^\ast \, d\omega 
		= {1 \over 2\pi} \int_{-\pi}^{\pi} {\tilde{u}(e^{i(\omega+\varphi)})\over 1+ \tilde{u}(e^{i(\omega+\varphi)})} {1+\tilde{u}(e^{i\omega}) \over \tilde{u}(e^{i\omega})} \, d\omega
		,
\end{eqnarray}
where we used Eq. (\ref{eqSM02}) 
.
To compute the integral, 
we set  
$e^{i\omega}\to z$, 
and with the useful factorization Eq. (\ref{eqSM05}), 
Eq. (\ref{fr}) becomes
%
%
\begin{equation}
	\tilde{F}(\varphi) 
		= 
			{1\over 2\pi i} \oint_{|z|=1} 
			\underbrace{ 
						{1\over z} \,
						{
						e^{i\varphi}
						\prod_{j=1}^{w-1} 
						(z_j - ze^{i\varphi}) 
						\over 
						\prod_{j=1}^{w-1} 
						(1-z_j^\ast ze^{i\varphi})
						} \,
						{
						\prod_{j=1}^{w-1} (1-z_j^\ast z) 
						\over 
						\prod_{j=1}^{w-1} (z_j - z)
						}
						}
						_{{\cal I}(z)} \, dz.
\end{equation}
Since $|z_j|<1$, the residues of the integrand ${\cal I}(z)$ are
$\text{Res}[{\cal I}(z), z_0] = e^{i\varphi}$ (where the trivial pole $z_0 = 0$) 
and $\text{Res}[{\cal I}(z), z_k] = - {e^{i\varphi}} z_k^{-1} \prod_{j=1}^{w-1}{(z_j - z_k e^{i\varphi}) (1-z_j^\ast z_k)\over (1-z_j^\ast z_k e^{i\varphi})}\,\prod_{j=1,j\neq k}^{w-1} {(z_j - z_k)^{-1}}$ for $ 1 \le k\le w-1$, 
which give a rather formal result
\begin{eqnarray}\label{eq13}
	\tilde{F}(\varphi)
							&=& \sum_{k=0}^{w-1} \text{Res}[{\cal I}(z), z_k] 
							= 
								e^{i\varphi} 
								- e^{i\varphi}(1-e^{i\varphi})
								\sum_{k=1}^{w-1}
								\bigg[
									\prod_{j=1}^{w-1} 
										{1-z_j^\ast z_k
										\over 
										1-z_j^\ast z_k e^{i\varphi}}
								\bigg]
									\prod_{j=1,j\neq k}^{w-1} 
									{z_j - z_k e^{i\varphi} 
									\over 
									z_j - z_k}.
\end{eqnarray}
\end{widetext}
Recall that 
\begin{equation}\label{eq14}
\begin{aligned}
	\tilde{F}(\varphi) 
						&= \sum_{n=1}^\infty \exp(in\varphi) \, F_n = \sum_{n=1}^\infty (1+in\varphi-n^2\varphi^2/2+\cdots) \, F_n \\
						&= P_\text{det} + i \varphi \expval{n}P_\text{det} - \varphi^2  \expval{n^2}P_\text{det}/2
						+ \cdots
						,
\end{aligned}
\end{equation}
where $P_\text{det}=\sum_{n=1}^\infty F_n$ is the detection probability, 
appearing here since the expectation values are actually conditional moments:  
$\expval{n^m} := \sum_{n=1}^\infty n^m F_n/P_\text{det}$,
and $P_\text{det}=1$ in our case study.
Expanding Eq. ({\ref{eq13}}) around $\varphi=0$ yields
\begin{equation}\label{eq15}
	\tilde{F}(\varphi) 
						= 1 + i \varphi w + \varphi^2  
						\bigg[
							-{w^2 \over 2}-\sum_{i,j=1} ^{w} {z_i z_j ^{*} \over 1 - z_i z_j^{*} }
						\bigg] 
						+ \cdots.
\end{equation}
A comparison between Eq. (\ref{eq14}) and Eq. (\ref{eq15}) shows
\cite{Gruenbaum2013} 
%
\begin{equation}
\begin{aligned}
\langle n \rangle = & \,\, w, \\
\mbox{Var}(n) = & \expval{n^2} - \expval{n}^2
			  	= \sum_{i,j=1} ^{w} 
			  	{
			  		2 z_i z_j ^{*} 
			  		\over 
			  		1 - z_i z_j^{*} 
			  	}.
\end{aligned}
\label{eq02}
\end{equation}
%
First of all, 
the mean of $n$ is quantized, namely, 
is equal to the number of the zeros of $\tilde{\phi}(z)$ 
or as mentioned the number of distinct energy phases corresponding to non-zero $p_k$. 
Further, $w$ has a topological meaning \cite{Gruenbaum2013} 
as it is the winding number of the generating function of $\phi_n$, see Appendix \ref{appA}.
Second, to compute the variance, one needs to get the zeros of $\tilde{\phi}(z)$
and then use the second equation in Eq. (\ref{eq02}). 
Via the procedures above, 
we can regard Eq. (\ref{eq13}) as a ``moment-generating function'' of $n$,
which in principle gives $\expval{n^k}$ in terms of zeros $\{z_i\}$ 
by merely expanding to an order $\varphi^k$, 
namely, $\expval{n^k} = (-i)^k \tilde{F}^{(k)}(0)/\tilde{F}(0)$ 
with $\tilde{F}(0) = 1$ the normalization.
Notice that if one or more zeros approach the unit circle $|z_i| \to 1$
we obtain a large variance and this will be the topic of our work.

\begin{figure*} 
\centering
\includegraphics[width=0.75\linewidth]{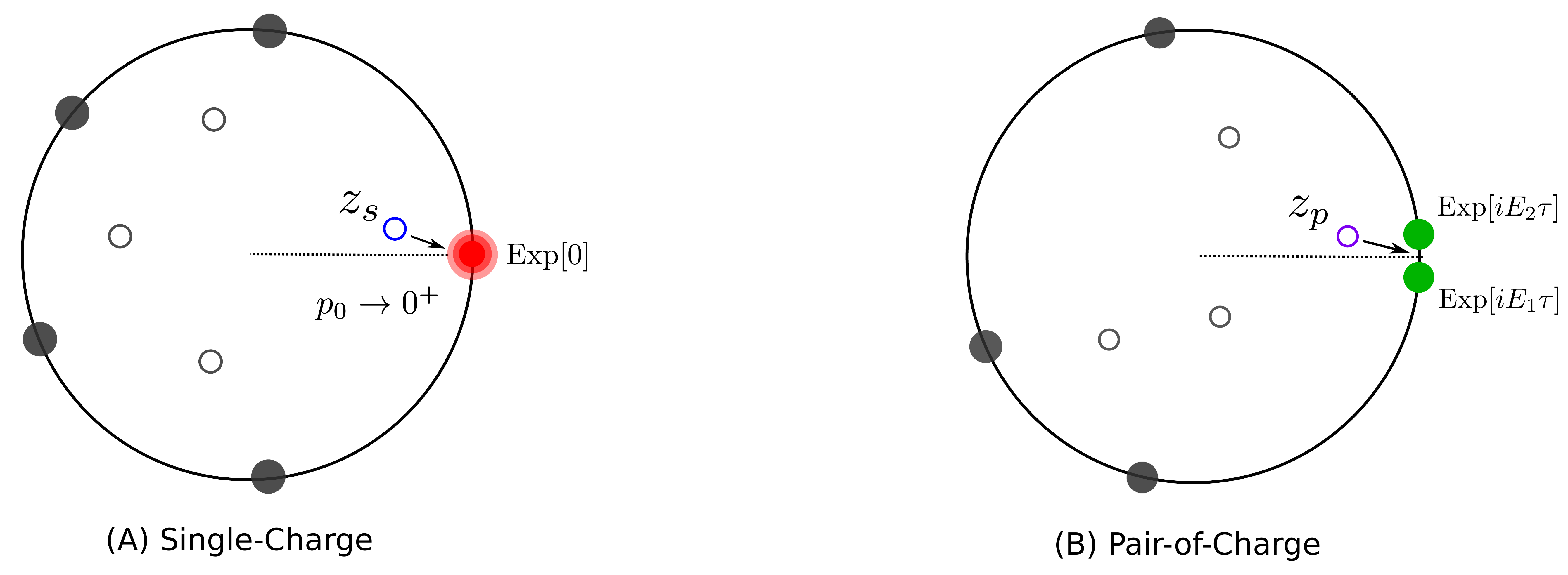}
\caption{(Color online)
A system with five distinct phases $e^{iE_j\tau}$.
These phases give the positions of positive charges on the unit circle
whose magnitudes are overlaps of energy states with the initial condition. 
In the unit disk we have four non-trivial zeros of the
corresponding  force field, 
these are on $\{ z_i\}$ here denoted with  empty circles \cite{Gruenbaum2013}.
Once we obtain the zeros, we have the quantum
fluctuations using Eq. (\ref{eq02}) 
while the mean is $\langle n \rangle = w = 5$. 
Here we show (A) the single-charge theory, where the red colored charge is weak,
hence a zero $z_s$ is found close to the unit circle (see arrow);
then (B) the two-charge theory, where the green colored charges are merging,
namely $\exp(iE_1\tau)\simeq \exp(iE_2\tau)$,
accordingly a zero $z_p$ is close to the unit circle (see arrow). 
Both in turn mean that we have large fluctuations in the system as the variance of $n$ Eq. (\ref{eq02}) will be large when $|z_s|\to 1$ or $|z_p|\to 1$.
}
\label{fig1}
\end{figure*}

We now discuss the $z_{i}$'s.
Generally finding the zeros can be difficult,
since these are roots of the polynomial ${\cal N}(z)$ Eq. (\ref{ND}) 
[or equivalently the zeros of $\tilde{\phi}(z)$].
Gr\"unbaum {\em et al} \cite{Gruenbaum2013} presented a mapping of the problem 
of finding the zeros $\{ z_i \}$ to a classical charge theory.
This allows us to find the zeros using our physical intuition from electrostatics.
From Eq. (\ref{eqSM02}) we see that $\{ z_i \}$ are solutions of
%
\begin{equation}\label{zerosol}
0  = \sum^w_{k=1} {p_k \over e^{iE_k\tau}-z} = {\tilde{u}(z) \over z} .
\end{equation}
Now the terms $p_k/(e^{iE_k\tau}-z)$ can be thought of as 
two-dimensional Coulomb forces,
generated by positive ``charges'' $p_k$ located on the unit circle, 
namely on the phase $\exp(iE_k\tau)$.   
The force field generated by the $w$ charges all on the unit circle is
\begin{equation}\label{forceF}
0 =  {\cal F}(z) 
= \sum^w_{k=1} {p_k \over e^{iE_k\tau}-z}.
\end{equation}
So we are searching for the stationary points of the classical force field.
Integration of ${\cal F}(z)$ with respect to $z$ reveals a two-dimensional Coulomb potential
$V(z) = \sum_{k=1} ^w p_k \ln|e^{i E_k \tau} - z|$.

To recap, in the electrostatic  picture shown schematically in 
Fig. \ref{fig1}, we have $w$ charges  located  on the unit circle, 
positioned at $\exp(i E_k \tau)$ and with magnitude $p_k$.  
In the unit disk we find $w-1$ stationary points of the force
field located on the zeros $\{z_i\}$ 
and $\expval{n} = w$.  
Once we find these stationary points, i.e. the $\{ z_i \}$  of the force field 
we can use Eq. (\ref{eq02}) to obtain the variance. 
When a zero approaches the unit circle, 
the fluctuations of $n$ are large 
because the denominator of the variance formula in Eq. (\ref{eq02}) vanishes \cite{Gruenbaum2013}.
The goal now is to find out explicitly the magnitude of these fluctuations,
and to better understand when the fluctuations are large.

~\\

\section{Asymptotic Formulas for the Variance of $n$}

\subsection{Single-charge theory}
Assume that one of the overlaps denoted
$p_0$, associated to energy $E_0$, is small $p_0\ll1$ 
and in particular much smaller than all the others. 
We will later present simple Hamiltonians which exhibit this property.  
In the electrostatic language we have a weak charge on $\exp( i E_0 \tau)$. 
We put $E_0=0$. Clearly, as shown in Fig. \ref{fig1}(A), we find a zero close to this charge, 
denoted $z_s\simeq 1$ with $s$ standing for single. 
On $z_s$, the electrostatic force vanishes, because the force of the weak charge balances all other forces
(in analogy, the equilibrium point in the sun-earth system is much
closer to earth than to the sun).
It follows that for this single-charge scenario, we have 
\begin{equation}
\mbox{Var}(n)\sim 
{2 |z_s|^2 \over  1 - |z_s|^2 },\;p_0\to0.
\label{eq02a}
\end{equation}
%
Using Eq. (\ref{forceF}) with perturbation theory, presented in Appendix \ref{b1}, we find
%
\begin{equation}
z_s \sim  1 - { p_0 \over  \sum_{j\neq 0} p_j/ [1-\exp(i E_j \tau)]},
\label{eq03}
\end{equation}
%
hence
\begin{equation}
\mbox{Var}(n) \sim {1 \over 2 p_{0}} \left\{1 + \left[ \sum_{j\neq 0} p_j \cot\left[ \left( {E}_j -{E}_0 \right)  \tau /2\right] \right]^2 \right\}.
\label{eq04}
\end{equation}
%
This is the first main result of this paper.
As expected the variance depends on all charges $p_{j}$.
The blow-up of the variance is easy to understand from the classical picture, but
what is the physics in the quantum problem? 
Roughly speaking, the process of repeated measurements may drive the system into a state that has 
considerable overlap with $\ket{E_0}$. Nevertheless, since $p_0$ is small, the particle is not 
efficiently detected. A typical outcome of $n$ may be much larger than $\langle n \rangle$ ($=w$), which implies large fluctuations of $n$ for small $p_0$, given by Eq. (\ref{eq04}). 
Notice that 
when $|E_j - E_0| \tau \simeq 2 \pi k$ for some $k$, 
we get a large contribution from the $\cot(\cdots)$ in Eq. (\ref{eq04}) (see Remark \ref{rem1}). 
In the electrostatic picture this is because two phases (and hence two charges) are merging 
and this is the topic of the next section, where we do not assume that $p_0$ is small. 
~\\

%

\subsection{Pair of charges}
\begin{figure} 
\centering
\includegraphics[width=\columnwidth]{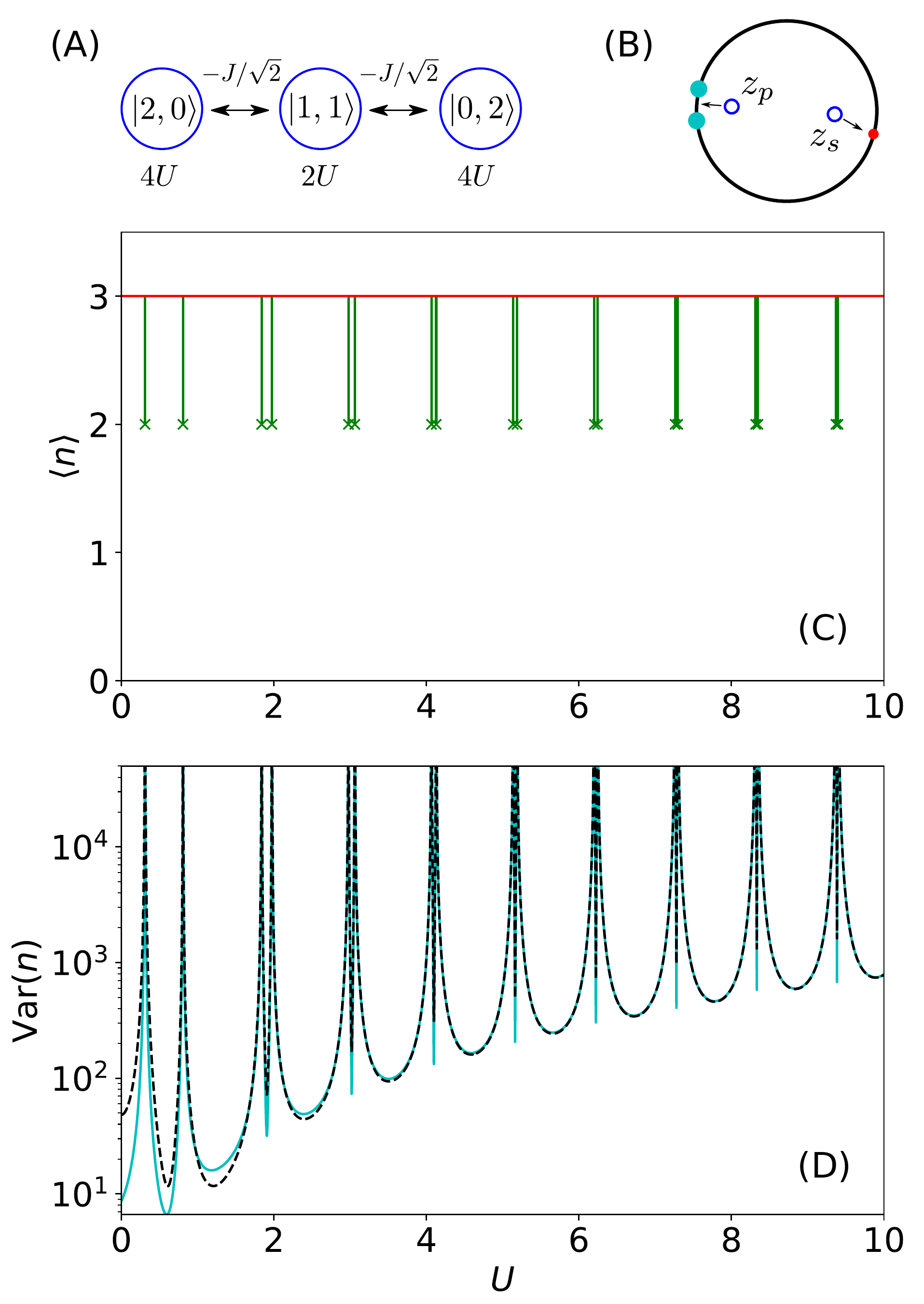}
\caption{(Color online)
(A) Interacting two-boson model schematics, (B) charge configurations,
(C) the mean of $n$ and (D) the fluctuations  of $n$ 
as a function of the interaction strength $U$ with  $J=1$ and $\tau=3$. 
The red horizontal line in (C) presents $\langle n \rangle$ for non-exceptional $U$, 
and the green crosses are the reduced winding number found for special $U$'s.
The peaks in (D) correspond to the discontinuous reduced $w$ in (C),
i.e. whenever $w=2$ we get diverging variance.. 
As shown in the upper right corner (B), 
when  $U$ is large an overlap/charge (colored red)
becomes small corresponding to single-charge theory 
and hence the zero $z_s$ is approaching this charge. 
Simultaneously,
two phases corresponding to a pair of charges (blue) merge when $U\to \infty$, 
and this means that a second zero $z_p$ is also approaching the unit circle. 
The theory for $\mbox{Var}(n)$, 
i.e. the  sum of Eqs. (\ref{eq04},\ref{eq06}) perfectly matches exact results
when $U$ is large. 
}
\label{fig3}
\end{figure}

Another mechanism leading to the blow-up of the variance is the case
when two energy levels, denoted $E_1$ and $E_2$, satisfy the resonance condition  $\exp( i E_1\tau) \simeq \exp(i E_2 \tau)$ \cite{Gruenbaum2013,Friedman2017a}.
Note that this can be achieved by modifying $\tau$ or some other parameter of $H$.
When  $\exp( i E_1\tau) =\exp( i E_2 \tau)$ exactly,
the winding number is reduced by one, and in the vicinity of this jump in $w=\langle n \rangle$ 
we get large fluctuations. 
The jump in $\langle n \rangle$ is not directly measurable, 
since it is found only for an isolated value of the control parameter, say $\tau$ 
[see Fig. \ref{fig3}(C) for an example]. 
The investigation of the variance is thus crucial
as it presents the  signature for this transition in its vicinity.
In our case we  have two charges $p_1$ and $p_2$ close to each other,
both located on the unit circle.
So we expect to find a zero, denoted $z_p$ with $p$ standing for pair,
in their neighborhood.
This is because 
the point of zero force is largely determined by this pair.
In analogy, the equilibrium point between two neighboring stars
is determined to leading order by these and not by other distant stars. 
See Fig. \ref{fig1}(B) for this case.

Since $z_p\simeq 1$ but $|z_p|^2<1$ we have
from Eq. (\ref{eq02})
$\mbox{Var}(n) \sim  2 |z_p|^2/ (1 - |z_p|^2)$.
We need to find an approximation for $z_p$ as $\delta\to0$, where
$\delta = ({{E_{2}}\tau-{E_{1}}\tau})/2\;\; \mbox{mod} \,2\pi$.
At $\delta=0$ the two phases merge. 
As explained in Appendix \ref{b2}, a second order expansion of Eq. 
(\ref{forceF})
in $\delta$ yields
%
\begin{equation}
z_p  \sim 1 + i { p_1 -p_2 \over p_1 + p_2} \delta +
\left[ { 4 p_1 p_2 \over \left( p_1 + p_2\right)^3}
 \sum_{j\neq 1,2} { p_j \over 
e^{i E_j \tau} -1} -   {1 \over 2} \right] \delta^2 .
\label{eq05}
 \end{equation}
%
The leading order term is unity  because we choose
the zero energy as $(E_1+E_2)/2=0$.
The first correction term depends only on $p_1$ and $p_2$ as expected,
while the last term is already sensitive to all the other charges $p_j$ with $j\neq 1,2$.
Importantly the first-order term has no real part, 
unlike the single-charge theory case. Since we are actually interested in $|z_p|^2$ the expansion must be carried out to second order. Here enters a little magic:
using the normalization condition $\sum_j  p_j=1$ and $1/(\exp[i x] -1) =
-1/2 - i \mbox{cot}(x/2)/2$, we find $|z_p|^2$ and the variance is
\begin{equation}
\mbox{Var}(n) \sim 2 {\left( p_1 + p_2\right)^3 \over p_1 p_2} {1 \over 
\tau^2 \left( \bar{E}_2 - \bar{E}_1 \right)^2 }
\label{eq06}
\end{equation}
where $\bar{E}_j \tau = E_j \tau$ $\mbox{mod} \  2\pi$.
Surprisingly the background charges (all the $p_j$s, for $j\neq1,2$) 
cancel out in the final formula. The asymptotic variance is
not sensitive to their presence, in contrast with Eq. (\ref{eq04}). 
However these charges cannot be neglected
in the calculation, see Eq. (\ref{eq05}). 
To put this differently: 
had the charges not satisfied the normalization condition, the final result would be sensitive to all the charges. So in this sense we are dealing with a classical
charge theory, but with three important constraints: all charges are positive, their sum is one and they are on the unit circle.
More importantly Eq. (\ref{eq06})  exhibits the blow-ups of the variance close to resonances which can be controlled for example by varying $\tau$. 
We now demonstrate these results with two examples.
~\\

%
%

\subsection{Two interacting bosons in a Josephson junction}
Two particles can occupy two states, left and right, and are governed by the Hamiltonian
\begin{equation}\label{bosonsH}
  H = -{J\over 2}(\hat{a}_l^\dagger \hat{a}_r + \hat{a}_r^\dagger \hat{a}_l) + {U}(\hat{n}_l^2 + \hat{n}_r^2),\;\;   \hat{n}_{l,r}=\hat{a}_{l,r}^\dagger \hat{a}_{l,r}
\end{equation}
where $\hat{a}_{l,r}^{\dagger}(\hat{a}_{l,r})$ is the creation (annihilation) operator on the left 
($l$) or right ($r$) well.
The Fock space is spanned by $\ket{2,0}$ and $\ket{0,2}$, 
i.e. both bosons on the left or right of the junction, and $\ket{1,1}$
one boson on each site.
This well-known system is described by tunneling elements $J$, 
and the interaction energy $U$, see sketch in Fig. \ref{fig3}(A)
and  further details in Appendix \ref{c1}.  
We start with two  particles on the left and investigate
the first return of this pair as we vary $U$, i.e., $|\psi_{\rm in}\rangle = |2,0\rangle$.
Since we have three distinct energy levels,
the winding number is $\langle n \rangle = w =3$ except for special values of $U$ 
where the variance of $n$ diverges, see Fig. \ref{fig3}(C).

In the limit of large $U$,
one of the overlaps/charges which we call $p_{0}$, 
becomes very small and the single-charge theory applies. 
Here, $p_0$ is the overlap of the detected state $\ket{\psi_{\rm in}}$ 
with the ground state $\ket{E_0}$. 
The vanishing overlap is understood easily. 
For large $U$, the ground state is almost
$\ket{1,1}$ 
and orthogonal to $\ket{\psi_{\rm in}}=\ket{2,0}$. 
More precisely, 
$p_{0} = |\langle 2,0|E_{0}\rangle|^2\sim J^2/8U^2 \ll1$. 
As we increase $U$ a second effect takes place, it is easy to show
that  two excited  energy levels approach each other
 $|E_1-E_2|\sim J^2/(2 U) \to 0$, so we get a contribution
also from the two-charge theory (see Appendix \ref{c1}). 
As shown in Fig. \ref{fig3}(B) the two effects imply two zeros
approaching the unit circle separately,  
and hence we can add up the two contributions Eq. (\ref{eq04}) and Eq. (\ref{eq06}) to reach excellent agreement between exact results and theory presented in the figure. 
As demonstrated in Fig. \ref{fig3} 
whenever we have a non-analytical jump $\langle n \rangle = 3 \to 2$ on isolated $U$ 
we find large fluctuations in the vicinity of the critical parameters. 
As mentioned since the discontinuous jump in $\langle n \rangle$ is found for isolated point 
(of measure zero) the measurements of the variance is the way to 
demonstrate the qualitative transition of the topological number $w$ in the system. 
~\\

\subsection{The ring}
\begin{figure*} 
\centering
    \includegraphics[width=0.85\linewidth]{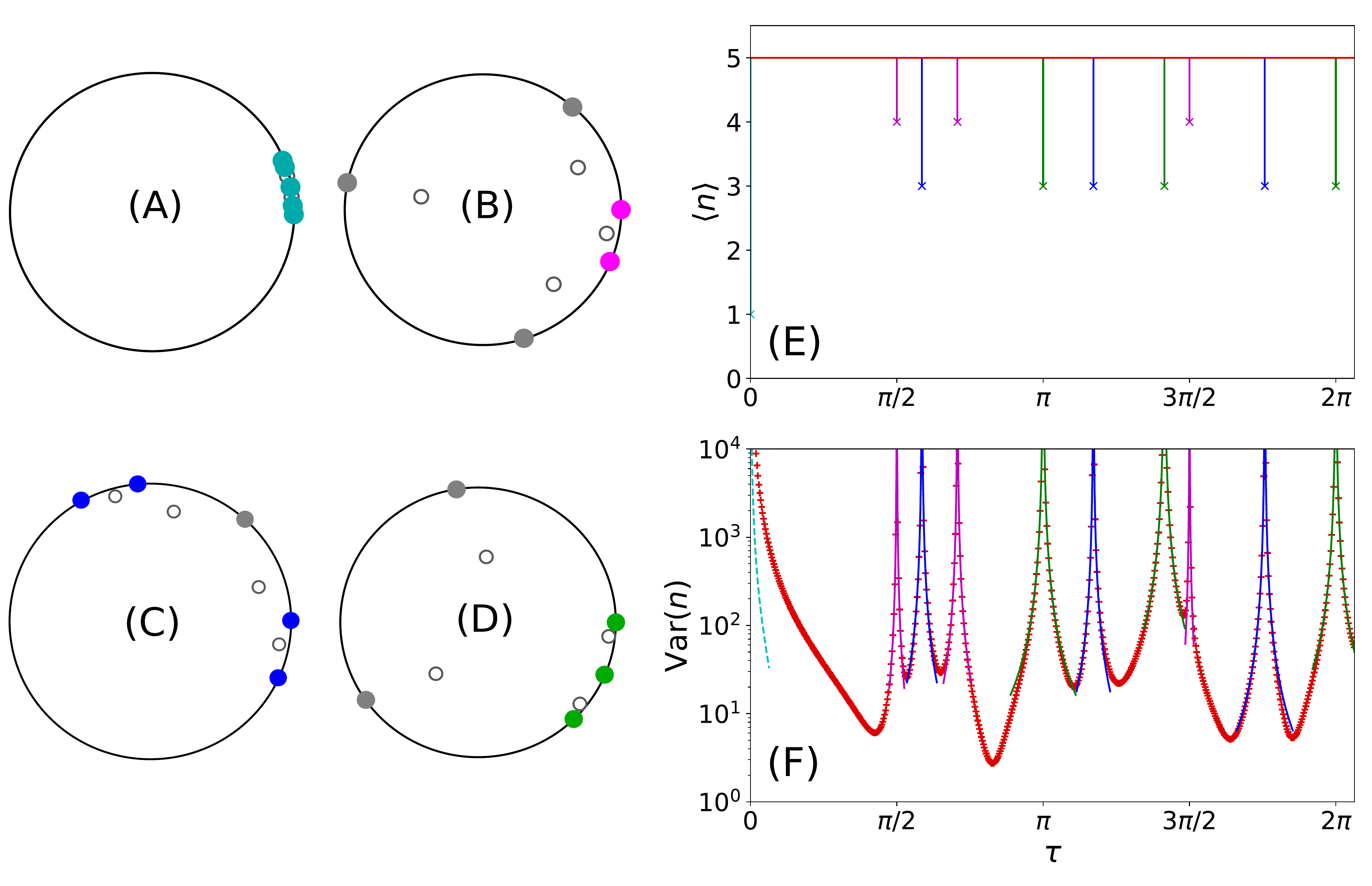}
\caption{(Color online)
(A-D) Charge configurations, 
(E) the mean of $n$ and 
(F) the variance of $n$ versus $\tau$ for the quantum return problem
on a ring of size eight with $\gamma=1$.  
As presented in (E), the winding number (red horizontal line) remains constant for general $\tau$. 
Then isolated jumps, shown by vertical lines with colors corresponding to scenarios (A-D), 
appear at some critical $\tau$,
around which
one finds diverging fluctuations of $n$ in (F).  
When the variance is large, corresponding to merging  phases,
we find rich physical behaviors which are captured by our theory: 
(A) the Zeno regime $\tau\to 0$ we have a five-charge theory described by the bound Eq. (\ref{eq10})
(dashed cyan line), 
(B) purple curves represent the two-charge theory Eq. (\ref{eq06}), 
(C) blue is the double two-charge theory, 
(D) green peaks are the triple-charge theory Eq. (\ref{eq09}).
Our formulas  perfectly match the exact results presented with the red curve. 
}
\label{fig2}
\end{figure*}

A nearest-neighbor tight-binding model on an eight-site ring has the Hamiltonian:
\begin{equation}
H = -\gamma \sum_{\bm{x}=0} ^7 \left[ \ket{\bm{x}} \bra{\bm{x}+1} + \ket{\bm{x}+1} \bra{\bm{x}} - 2 \ket{\bm{x}} \bra{\bm{x}} \right]. 
\label{eq07}
\end{equation}
where $\ket{0}=\ket{8}$, $\gamma$ is the hopping rate.
This Hamiltonian describes hopping between nearest neighbors.   
Consider a particle initially localized on a site of the ring
$\ket{\psi_{\rm in}} = \ket{0}$
and we investigate how
$\mbox{Var}(n)$ is controlled by the sampling time  $\tau$.
 In this model we have  five distinct  energy levels because of degeneracies 
(the energy levels are  $E_k/\gamma= 2 - 2\cos[\pi k/4]$ with $k=0, ..,7$). 
This means that except for the special sampling times
the winding number is $\langle n \rangle = w=5$. 
The exceptional sampling times are given by $\Delta E \tau= 2 \pi j$ 
where $\Delta E=|E_{k^\prime} -E_k|\tau$
for any pair of energies in the system. 
When $\tau$ approaches this limit, we find large fluctuations (see Fig. \ref{fig2}).

There are no effectively small charges, 
so we expect to find the scenario of the two-charge theory.
However, in reality the physics of this system (and of other simple examples)
is richer than what we have found so far.
As shown in Fig. \ref{fig2}
we have four categories. 
When $\tau \to 0$ we have the Zeno regime \cite{Facchi2008}, 
in this case all the five phases $\exp(i E_j \tau)$ converge to unity, 
so here in principle we must locate the stationary points arising from a configuration of five charges [see Fig. \ref{fig2}(A)].
We also have cases where three phases approach each other on the unit circle [Fig. \ref{fig2}(D)]. 
So our theory based on pair of charges or single charge is not sufficient 
and we will soon consider these interesting cases in some detail.
We also find cases where two pairs of charges  converge at different locations
on the unit circle [see Fig. \ref{fig2}(C)], 
here we may use our results and sum up the two contributions. 
Finally we have the cases where the two-charge theory holds. 
The comparison in Fig. \ref{fig2}(F) between our theory 
and an exact diagonalization of the problem shows excellent agreement.
Note that for the Zeno regime (five-charge theory),
we have plotted a lower bound to be discussed soon.  
Motivated by these observations we now extend further the basic theory, 
revealing two more mechanism for the blow-up of the fluctuations. 

%
%
~\\

\subsection{Triple-charge theory}
It is common that three (or more) phases merge on the 
unit circle, for example in systems with commensurate energy levels or
in the $\tau\to 0$ limit. 
The corresponding zeros $\{z_i\}$ may exhibit interference effects as the off-diagonal terms in 
Eq. (\ref{eq02}) may become important.
Consider the case where three phases are close by on the unit circle.
Here, we will consider a symmetrical situation to reduce the number of free parameters.
The three energy levels are $E_0=0$ and $E_{\pm}=\pm E$ with phases
$e^0$ and $e^{\pm iE\tau} =e^{\pm i 2 \pi  k \pm i\delta}$ where
$k$ is an integer and $\delta$ is our small parameter. 
This configuration of charges/phases yields two zeros called $z^{\pm}_d$, 
and $d$ stands for dimer.  
These are located in the vicinity of the unit circle, 
as expected from basic electrostatics, see Fig. \ref{fig2}(D).
We denote $p_0=|\bra{\psi_{{\rm in}}} E_0 \rangle|^2$ and
$p=|\bra{\psi_{{\rm in}}} E_{\pm} \rangle|^2$, 
this corresponds to the example of the ring.
Using Eq. (\ref{eq02}) in this case we have 
\begin{equation}
\mbox{Var}(n) \sim \sum_{\sigma= \pm}
{2 |z_{d} ^{\sigma }|^2  \over 1 - |z_{d} ^{\sigma }|^2}
+ 
\underbrace{ 
\left( 
 {2 z_{d} ^{+} (z_{d} ^{-})^* \over
 1-  z_{d} ^{+} (z_{d} ^{-})^* } + c.c.
\right).}
_{\cal M} 
\label{eq08}
\end{equation}
%
From the example's symmetry
we have $z_{d} ^{+} = (z_{d} ^{-})^*$.  
A detailed calculation, presented in Appendix \ref{b3} shows that as long as $p$ is not small, 
\begin{equation}
\mbox{Var}(n) \sim 16 { \left( p_0 + 2 p \right)^2 \over p} 
{1 \over \tau^2 \left( \bar{E}_{+} - \bar{E}_{-} \right)^2}. 
\label{eq09}
\end{equation}
Again, this describes gigantic fluctuations as the three phases are merging 
and perfectly matches those peaks in Fig. \ref{fig2}(F).
The mixed terms become negligible in the limit,
more specifically as shown in Appendix \ref{b3}
$\lim_{\delta\to 0} {\cal M} = -2 + p/[p_0(p_0+2 p)]$. 
 The physical  reason is that
in the limit the two zeros overlap and then they cannot interfere. 
~\\

\begin{remark}\label{rem1}
Note that the former three formulas Eqs. (\ref{eq04}, \ref{eq06}, \ref{eq09}) are correlated.
In Eq. (\ref{eq04}),
assuming some $l$, $(E_l \tau-E_0)\tau \simeq 2\pi k$ with $k$ an integer, namely, $(\bar{E}_l \tau-\bar{E}_0)\tau=\delta$ with $\delta\to0$, using $\cot(\delta/2)\sim2/\delta$, we obtain $\text{Var}(n)\sim 2p_l^2/[p_0(\bar{E_l}-\bar{E_0})^2\tau^2]$.
Then recalling Eq. (\ref{eq06}), if $p_l\ll 1$ and $p_{l'}$ finite, 
further simplification gives the same expression.
As to Eq. (\ref{eq09}), if the upper and lower charges $p$ are small but $p_0$ is finite, 
we get $\text{Var}(n) \sim 16 p_0^2/[p\cdot (\bar{E}_+ - \bar{E}_-)^2\tau^2]$. 
This is consistent with the single-charge theory Eq. (\ref{eq04}) but with two ``single-charge'', namely we have two $z_s$ [see Eq. (\ref{eq03})]. 
Since we have the middle charge close by the two charges with small $p$, 
applying the same approximation of $\cot(\dots)$, 
we get $\text{Var}(n)\sim 2\cdot 2p_0^2/[p \cdot (\bar{E}_0-\bar{E}_+)^2\tau^2]$, 
here $(\bar{E}_0-\bar{E}_+)\tau$ is half of $(\bar{E}_+ - \bar{E}_-)\tau$, 
thus it is consistent with the single-charge theory.
\end{remark}
~\\

\subsection{Symmetry breaking in a ring system with defect}
%
\begin{figure}[htbp]
\centering
\includegraphics[width=0.8\columnwidth]{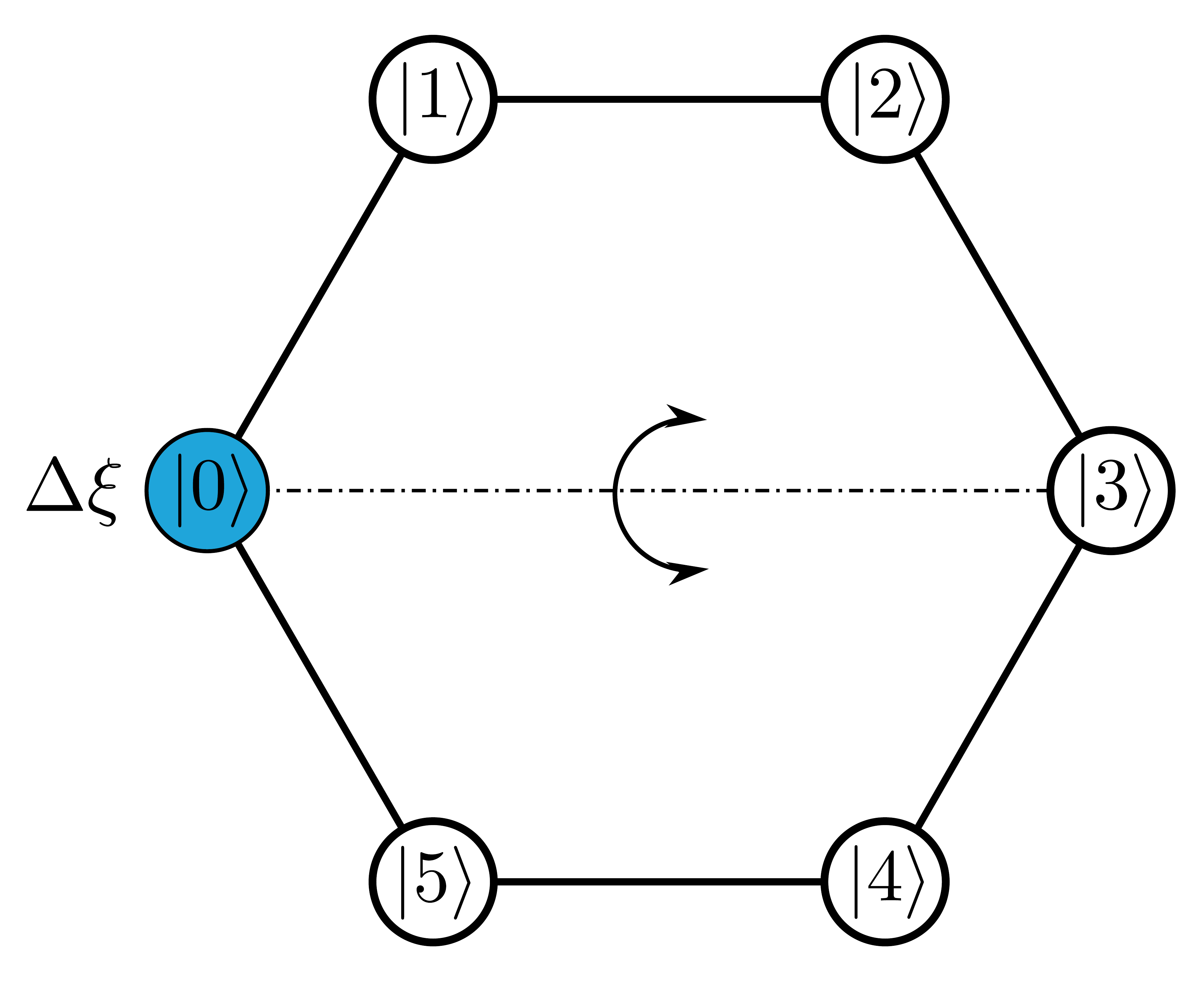}
\caption{(Color online)
Schematic sketch of the six-site ring with one defect.
The circle colored blue shows where the defect is.
The introduction of defect $\Delta\xi$ breaks the original rotational invariance, 
resulting in reflection symmetry (see the dashed line).
}
\label{fig6s}
\end{figure}
\begin{figure*}[htbp]
\centering
\includegraphics[width=1.0\linewidth]{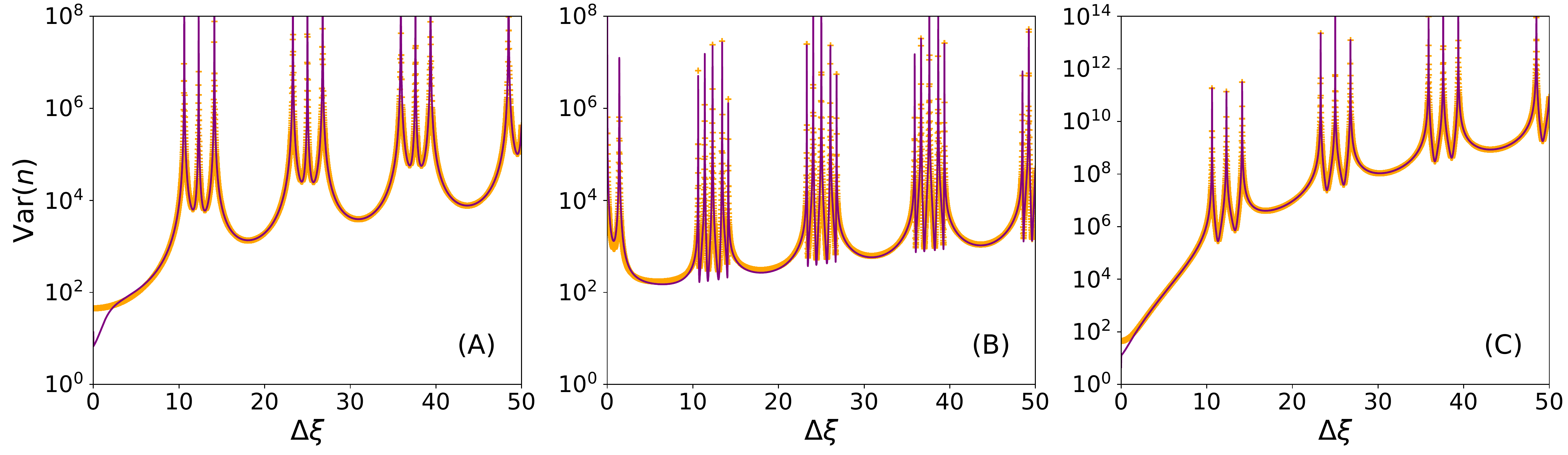}
\caption{(Color online)
Variance of $n$ for six-site ring model with one defect.
Here $\gamma$ is set to $1$, and the defect strength $\Delta\xi$ ranges from $0$ to $50$.
(A) when the detector set at site $x=0$, 
where the defect is located;
(B) when the detected/initial state is $\ket*{1}$, the neighbor of the defect;
(C) when the state $\ket*{3}$ is the target.
The figures show perfect matching 
between our theory (purple curves) and exact results (orange crosses).
Note that in this system, the quantum fluctuations are gigantic even far from resonances.
In case (B) the measurements break the reflection symmetry of the system, 
making the qualitative features of statistics of returns different 
from the measurements maintaining the symmetry.
See further details in the text.
}
\label{fig6d}
\end{figure*}
We now consider a ring with six sites as our final example. 
One site denoted $\ket*{0}$ is defected in the sense that its on-site energy is $\Delta\xi$, 
while other on-site energies are zero. 
Hopping is between neighboring sites similar to previous example.  
The schematics of our system is shown in Fig. \ref{fig6s}.
The Hamiltonian reads
\begin{equation}
H = -\gamma \sum_{\bm{x}=0}^5 \left[ \ket{\bm{x}} \bra{\bm{x}+1} + \ket{\bm{x}+1} \bra{\bm{x}} - \Delta\xi \ket{{0}} \bra{{0}} \right],
\label{eq1}
\end{equation}
where $\ket*{{6}} = \ket*{{0}}$, satisfying the periodical boundary condition, and $\Delta\xi > 0$.
In this system we break the symmetry of $H$ when $\Delta\xi\neq 0$. 
We now address how this influences return time statistics.

We consider the return problem for three cases, 
choosing $\ket{\psi_{\rm in}} = \ket{0}$, $\ket{1}$ and $\ket{3}$. 
Importantly, when we measure on $\ket{0}$ and $\ket{3}$ we do not break the reflection symmetry of the system,
while when we measure on $\ket{1}$ we do. 
First let us consider the mean $\langle n \rangle= w$.
The first thing to do is to search for the energy levels of the system. 
While when $\Delta\xi = 0$ we have four energy levels; 
the perturbation breaks symmetry and now we have six distinct energy levels 
(when the splitting is small corresponding to weak $\Delta\xi$ we expect two-charge theory to hold, 
as degenerate energy levels split into two). 
When $\Delta\xi$ is large we find a state very much localized on the defect, 
which suggests that weak-charge theory is expected to play a role. 
While these effects are certainly present, a more subtle feature is related to symmetry.
Solving for the energy states we find that two eigenstates of $H$ are:
$\{0, 1/2, -1/2, 0, 1/2, -1/2 \}$, $\{0, -1/2, -1/2, 0, 1/2, 1/2\}$, 
corresponding to eigenvalues $\gamma, -\gamma$ respectively.
See Appendix \ref{c3} for further details. 
Now recall that the charges are the overlaps of the energy eigenstates and the detected state. 
From this we see that if the detection is on site $\ket{0}$ or $\ket{3}$ 
two charges will have zero magnitude. 
The consequence is that
\begin{equation}\label{symbw}
 w = 
\begin{cases}
	4, & \text{if $\ket{\psi_{\rm in}} = \ket{0}$ or $\ket{3}$} \\
    6, & \text{if $\ket{\psi_{\rm in}} = \ket{\bm{x}}$ with $\bm{x} \neq 0,3$}
\end{cases}
	.
\end{equation}
As explained, this result is valid as long as the phases of the system 
$\exp(i E_k \tau) \neq \exp(i E_{k'}\tau)$ do not overlap. 
When this happen we get a blow-up of the variance presented in Fig. \ref{fig6d}. 

The variance of $n$ for the two classes of initial conditions exhibits qualitatively different behavior.
When detection does not break symmetry we have clusters of three peaks, 
and for symmetry breaking measurement we have five, i.e. when $\ket{\psi_{\rm in}} = \ket{1}$, 
see Fig. \ref{fig6d}.
This is related to the fact that 
when symmetry is preserved the effective dimension of the Hilbert space is reduced ($w=4$) 
if compared with the symmetry breaking case ($w=6$). 
An interesting effect for $\text{Var}(n)$ is found for weak perturbations
$\Delta\xi \ll 1$. 
As shown in Fig. \ref{fig6d}(B) the fluctuations blow up in this limit 
when we break the symmetry i.e. $\ket{\psi_{\rm in}} = \ket{1}$.
The opposite is found when symmetry is maintained, 
Figs. \ref{fig6d}(A,C): now there is no blow-up of fluctuations.
To explain this, note that when symmetry is not broken,
$w$ for $\Delta\xi=0$ is a constant equal four and it remains four also for $\Delta \xi> 0$.
In contrast, when we break symmetry, we have a transition from $w=4$ for $\Delta\xi=0$,
to $w=6$ when $\Delta \xi$ is small (due to removal of degeneracies, and non-zero overlaps).
Since large fluctuations are found when $w$ performs a discontinuous jump,
we find diverging fluctuations only when the detected state breaks symmetry.
Our theory works nearly perfectly 
and thus in Fig. \ref{fig6d} it is hard to distinguish between predictions and exact solutions. 
The details of the theory and classification into weak-charge, two-charge theory etc., 
are provided in the Appendix \ref{c3}.  
~\\

\subsection{Zeno regime}
As we increase the number of merging charges, the calculation of the variance
becomes exceedingly hard. Such a case is the Zeno regime,
$\tau\to0$,
when all phases $\exp( i E_k \tau)$ coalesce. 
Basic electrostatics tells us that
all zeros are located in the convex hull of the charges \cite{Gruenbaum2013},
the area of which vanishes as $\tau\to 0$ (see schematic diagram in Appendix \ref{b4}). 
We may use this to our advantage and obtain a lower bound using basic geometry 
(see Appendix \ref{b4} for further details)
\begin{equation}
	\mbox{Var}(n) 	\ge 
						\left(  w - 1  \right) 
						\left[ 
								2 \cot^2\left( \Delta E_\text{m} \tau /2 \right) - w +2 
						\right],
\label{eq10}
\end{equation}
where $\Delta E_\text{m} = E_{{\rm max}} - E_{{\rm min}}$ is the width of the energy spectrum.
This useful bound shows that the variance diverges as $\tau\to 0$. It is plotted
in Fig. \ref{fig2}(F) for the ring example. 
~\\

\subsection{Time-energy uncertainty}
In the right-hand side of Eq. (\ref{eq10}), 
we can further simplify $\cot(\cdots)$ by using $\cot(x) \sim 1/x$ with $x \ll 1$, 
then we obtain a uncertainty-like relation of time and energy: 
\begin{equation}\label{UncertRe}
	(\Delta E_\text{m})^2 (\Delta t_\text{det})^2 \gtrsim 8 (w-1) \hbar^2,
\end{equation}
where $\Delta t_\text{det} $ is the standard deviation of the first detected return time, 
which is defined as
$\Delta t_\text{det} = \sqrt{\expval{(n\tau)^2} - \expval{n\tau}^2} = \tau \sqrt{\text{Var}(n)}$.
Here we recover the $\hbar$ to formulate the relation Eq. (\ref{UncertRe}). 
In the derivation we assume 
$\Delta E_\text{m} \tau \ll \hbar$.
In the uncertainty principle we relate
the variance of the first detected return time $t_{\rm det}$, 
the width of the energy spectrum, the topological number $w$ 
and the Planck constant $\hbar$. 
In the mathematical limit $w=1$, there is only one energy level, 
then $\Delta t_\text{det} = 0$ as the particle is detected at the first attempt, 
so $n=1$ with probability one for any sampling time $\tau$. 
Indeed the right-hand side of Eq. (\ref{UncertRe}) gives $0$ in this limit,
so the presence of the factor $w-1$ is physically reasonable.

\section{Discussion}

We provided a valuable theoretical background 
which gives explicit formulas for the large quantum fluctuations of the first detected return times.
The protocol of measurement relies on unitary evolution pierced by repeated measurements, 
a theme that is now used in many theoretical works. 
As the interest in quantum information and monitoring increases, 
a deeper understanding of the measurements' backfire 
and its influence on the dynamics become essential.  
Our work shows how this combination can yield diverging fluctuations. 

With the development of experimental technique,
it is now possible to perform single-atom quantum walks experiments in laboratory, 
e.g. in Refs. \cite{karski2009quantum,sherson2010single}, 
and single-atom imaging has been perfectly achieved with single-site resolution. 
In Ref. \cite{sherson2010single} the same model as our first example, 
Bose-Hubbard model, was employed in their experimental research with a number of particles.
And the investigation of this boson model involved Mott insulating state, 
a phase in which the on-site interacting strength $U$ is much greater 
than the tunneling element $J$.
This is also discussed in this paper, 
with smaller scale of two wells and a pair of bosons, 
namely a few-particle system rather than a many-body system.
Recent works \cite{Nitsche,xu2018measuring,Cohen2019,PengXue2019} indicates 
growing awareness in repeated measurements piercing unitary evolution, 
such experiments being based on single photon or coherent laser pulses.
The ring model with six sites has been implemented in Ref. \cite{Cohen2019}. 
Ref. \cite{Nitsche} nicely demonstrated the effects of repeated local measurements on unitary dynamics,
similar to the projective measurement under study here.
Therefore, with the techniques of single-atom \& single-site imaging 
or single-photon/coherent-laser-pulse platforms, 
or even other approaches mentioned in Ref. \cite{JingBo2013},
the theory developed here could be tested via these
discrete-time quantum dynamics setups. 


We hope that others will elaborate more on this theme, 
as many open questions remain to be explored. 
For example:  are the observed effects discussed here relevant to other measurement protocols? 
In the limit of small sampling time $\tau$, Dhar {\em et al} \cite{Dhar2015} showed that
the repeated measurement protocol is equivalent to the well known non-Hermitian \cite{moiseyev2011non} description of leaking, i.e. non-norm-conserving systems. 
The latter approach is used extensively in quantum optics \cite{Knight1998}, 
where the radiative decay rate plays the role of $1/\tau$, 
while in other systems the non-Hermitian dynamics is related to a sink term, 
e.g. in light-harvesting systems \cite{Caruso2009}.   
Our preliminary results in this direction show that 
the Zeno limit of the current work is well suited to describe non-Hermitian dynamics. 
Secondly what is the effect of disorder? 
If the disorder is static, localization plays an important role.
Then overlaps of energy states with localized measurement are typically small,
leading to what we termed ``small charge'' (however here we have more than one small charges). 
In this case the zeros $z_i$ are random, and further work is needed to evaluate their distribution. However, the basic formalism, especially the weak-charge limit, 
is expected to be of value also in this direction. 
Many other topics remain to be explored,
like the effects of weak measurements \cite{Brun2002} 
or coupling to the environment \cite{Gefen2005,Gherardini2018,Saito2018} 
or random sampling times \cite{Krovi2008} on the quantum return time.
The last one could modify the effects discussed here within the two \& three-charge theories, 
since these effects are related to partial revivals of the wave packet.
In contrast the physics of weak charges is expected to be more robust to changes 
in the protocol of measurement, 
since weak overlaps imply small probability of recording the particle.

To summarize,
our classification describes rich physics.
Based on the stroboscopic measurement protocol, 
the investigation of the statistics of first detected return times shows that
the mean $\langle n \rangle$ is equal to the winding number $w$ \cite{Gruenbaum2013},
its value fixed apart from exceptional parameter values.
In contrast, the variance exhibits rich physical behaviors.
It diverges whenever the winding number changes.
Our theory is based on four layers: 
single-charge theory where the variance is sensitive to the background charges  Eq. (\ref{eq04}),
two-charge theory describing the variance when two phases merge on the unit circle 
and the variance is insensitive to the background charges Eq. (\ref{eq06}), 
triple-charge theory where a dimer is found to describe two zeros 
in the vicinity of the unit circle Eq. (\ref{eq09}),
and finally a bound for the Zeno regime Eq. (\ref{eq10}), a many-charge scenario, 
in which we also provide a time-energy uncertainty relation Eq. (\ref{UncertRe}). 
The latter exploits the fact that the return time is fluctuating, 
while in the standard uncertainty relation $\Delta E \Delta t > \hbar$, $\Delta t$ is actually fixed.
Further it provides a relation that relates uncertainty to the topology,
i.e. the winding number $w$ enters in our relation. 

\begin{acknowledgments}
EB thanks P. H\"anggi and D. Kessler for valuable discussion. 
The support of Israel Science Foundation grant no. 1898/17 
and the Humboldt foundation is acknowledged. 
FT is supported by DFG (Germany) under grant no. TH 2192/1-1 and
KZ by the Julian Schwinger Foundation.
\end{acknowledgments}

\appendix

\section{Winding number}\label{appA}

Here we provide a topological interpretation for the mean of $n$: winding number, 
and an alternative derivation for the quantization of $\langle n \rangle$.
As mentioned above, the mean of $n$ is equal to the winding number of $\tilde{\phi}(e^{i\omega})$ with $-\pi\le\omega\le\pi$:
\begin{equation}
\begin{aligned}
\langle n \rangle 	&= (-i)\partial_\varphi \tilde{F}(\varphi)|_{\varphi=0} 
					= \bigg[
							{1 \over 2\pi i} \int_{-\pi}^\pi \partial_\varphi {\tilde{\phi}(e^{i(\omega+\varphi)}) \over \tilde{\phi}(e^{i\omega})} \, d\omega 
						\bigg]_{\varphi=0}\\
					&= \bigg[
							{1 \over 2\pi i} \int_{-\pi}^\pi \partial_\varphi e^{i[f(\omega+\varphi)-f(\omega)]} \, d\omega
						\bigg]_{\varphi=0} 
					= {1 \over 2\pi} \int^\pi_{-\pi} 
					{\partial_\omega} f \, d\omega.
\end{aligned}
\label{winding}
\end{equation}
Here we use the property $|\tilde{\phi}(e^{i\omega})|=1$ so that we can write as $\tilde{\phi}(e^{i\omega})=e^{if(\omega)}$. 
This equation represents ``winding'' behaviors of 
the generating function with $z$ on the unit circle in the complex plane. From the spectral decomposition Eq. (\ref{eqSM02}), using the identity $1/[\exp(ix)-1]=[-1-i\cot(x/2)]/2$, 
we get
\begin{equation}
f(\omega) = 2 \, \text{ArcTan} \bigg[ \sum_k^w p_k \cot {E_k\tau-\omega \over 2} \bigg].
\end{equation}
We use below the standard domain of the principle value of $-\pi/2<\text{ArcTan}(x)<\pi/2$.
Note that there exist a unique $\omega_k = \bar{E}_k\tau$ such that
$\cot [( E_k \tau \mp \epsilon -\omega_k )/2] \to \mp \infty$ for $\epsilon\to 0^{+}$. 
Using $\text{ArcTan}[\pm \infty] = \pm \pi/2$ we notice that in the small vicinity of $\omega_k=\bar{E}_k\tau$ 
\begin{equation}
\lim_{\epsilon \to 0} [f(\bar{E}_k\tau + \epsilon) - f(\bar{E}_k\tau - \epsilon)] = - 2 \pi
\label{eqWIN13}
\end{equation}
%
provided that $p_k\neq 0$.
We see that on $\omega_k$, $f(\omega)$ experiences a jump of size $2 \pi$,
so clearly in this representation $f(\omega)$ is not smooth in $[-\pi, \pi]$. 
Since in Eq. (\ref{winding}) we take a derivative of $f(\omega)$ 
the formula must be treated with some care.
We perform the integral by parts:
\begin{widetext}
\begin{equation}
\begin{aligned}
\langle n \rangle 	&= w =
						{1 \over 2 \pi} \lim_{\epsilon\to 0}
						\left[ 
						\int_{-\pi}^{\bar{E}_{1}\tau - \epsilon} \partial_\omega f d\omega
						+ \int_{\bar{E}_1\tau + \epsilon} ^{\bar{E}_2\tau -\epsilon} \partial_\omega f  d\omega  + \cdots \int_{\bar{E}_w\tau + \epsilon }^{\pi} \partial_\omega f d\omega 
						\right] \\
					&= {1 \over 2 \pi} \lim_{\epsilon \to 0}
					\left[
						f\left( \bar{E}_1\tau - \epsilon \right) - f(-\pi) + f\left(\bar{E}_2\tau -\epsilon \right) - f(\bar{E}_1\tau + \epsilon) + \cdots 
						+ f(\pi) - f(\bar{E}_w\tau + \epsilon) 
					\right] 
\label{eqWIN15}
\end{aligned}
\end{equation}
\end{widetext}
Note that $f(-\pi)-f( \pi)= 0$ and using Eq. (\ref{eqWIN13})
we see that $\langle n \rangle = w$. 
%
~\\



\section{Perturbation method}\label{appB}

We now determine the fluctuations of $n$ for the different classification schemes 
discussed in the text,
namely single-charge theory, pair-of-charge theory etc. 
Using Eq. (\ref{forceF}) or equivalently the first line in Eq. (\ref{ND})
the zeros are given by:
%
\begin{equation}
\sum_{k=1} ^{w} p_k \prod_{j\ne k} \left( e^{i E_j \tau} - z_i \right) = 0. 
\label{eqSM06}
\end{equation}
We will also apply perturbation method to this equation for simplicity in calculation.
~\\

\subsection{Single-charge theory}\label{b1}

This case, as mentioned in the main text, is set with one effective weak charge $p_0\ll 1$ the corresponding energy is  $E_0=0$.
From the force balance ${\cal F}(z)=0$, Eq. (\ref{forceF}) we find
\begin{equation}\label{px00}
0 = {p_0 \over 1-z} + \sideset{}{^\prime}\sum_k {p_k \over e^{iE_k\tau}-z},
\end{equation}
where $\Sigma^\prime$ means summation over all $k$ except for $k=0$. 
Assuming that $z_s\sim 1-\epsilon$, we find in leading order
\begin{equation}
0 = {p_0 \over \epsilon} - \sideset{}{^\prime}\sum_k {p_k \over 1-e^{iE_k\tau}},
\end{equation}
which yields 
\begin{equation}
\epsilon \sim {p_k \over \sideset{}{^\prime}\sum_k p_k/(1-e^{iE_k\tau})}. 
\end{equation}
Alternatively, we can plug the ansatz $z_s\sim 1-\epsilon$ into Eq. (\ref{eqSM06}) to obtain:
\begin{equation}
0 = p_0 - \epsilon \sideset{}{^\prime}\sum_k {p_k \over 1-e^{iE_k\tau}}
\end{equation}
which gives the same result.
%
From here the results Eqs. (\ref{eq03}, \ref{eq04}) follow. 
\\~

\subsection{Pair of charges}\label{b2}

When a pair of charges is nearly  merging,
 say $\exp(iE_1\tau)\simeq \exp(iE_2\tau)$
a zero denoted $z_p$ will approach the unit circle
in the vicinity of the charges, see schematic Fig. \ref{fig1}(B). 
For simplicity, as mentioned in the text, we assign the zero energy $(E_1+E_2)/2=0$, then rewrite $\exp(iE_1\tau)=\exp(-i\delta), \exp(iE_2\tau)=\exp(i\delta)$
and $\delta\ll 1$ is the small parameter of the problem.
Note that
    $2\delta = {\bar{E_{2}}\tau-\bar{E_{1}}\tau}$.
Inserting $z_p=1-C_1\delta-C_2\delta^2$ in Eq. (\ref{eqSM06}), 
regrouping $\delta,\delta^2$ terms, we find
\begin{widetext}
\begin{equation}\label{px3}
\begin{aligned}
    0 = \bigg[(p_1+p_{2})C_1 + i(p_1-p_{2}) \bigg]A_{1,2}\; \delta \; +
        \bigg\{ 
        \big[(p_1+p_{2})C_2 - {p_1+p_{2} \over 2}\big]A_{1,2} 
        + {\sum_{j\neq 1,2}p_j(1+C_1^2)B_{j,1,2}} 
        \bigg\} \; \delta^2
        + {\cal O}(\delta^3)
\end{aligned}
\end{equation}
%
%
where
\begin{equation}\label{allbll}
\begin{aligned}
    A_{1,2}&=\prod_{k\neq 1,2}(e^{iE_{k}\tau}-1),\,\,\,
    B_{j,1,2}&=\prod_{k\neq j,1,2}(e^{iE_{k}\tau}-1).
\end{aligned}
\end{equation}
Solving Eq. (\ref{px3}), we get
\begin{equation}\label{c1c2}
\begin{aligned}
    C_1 &= -i\;{p_1-p_{2} \over p_1 + p_{2}}, \,\,\,
    C_2 &= {1\over2} {- {4 p_1p_{2}\over (p_1+p_{2})^3}\sum_{j\neq 1,2}{p_j \over e^{iE_j\tau}-1}}.
\end{aligned}
\end{equation}
%
Then with some further  algebra we get Eqs. (\ref{eq05},\ref{eq06}). 
\end{widetext}
~\\

\subsection{Triple-charge theory}\label{b3}

\begin{figure}
\centering
\includegraphics[width=0.48\textwidth]{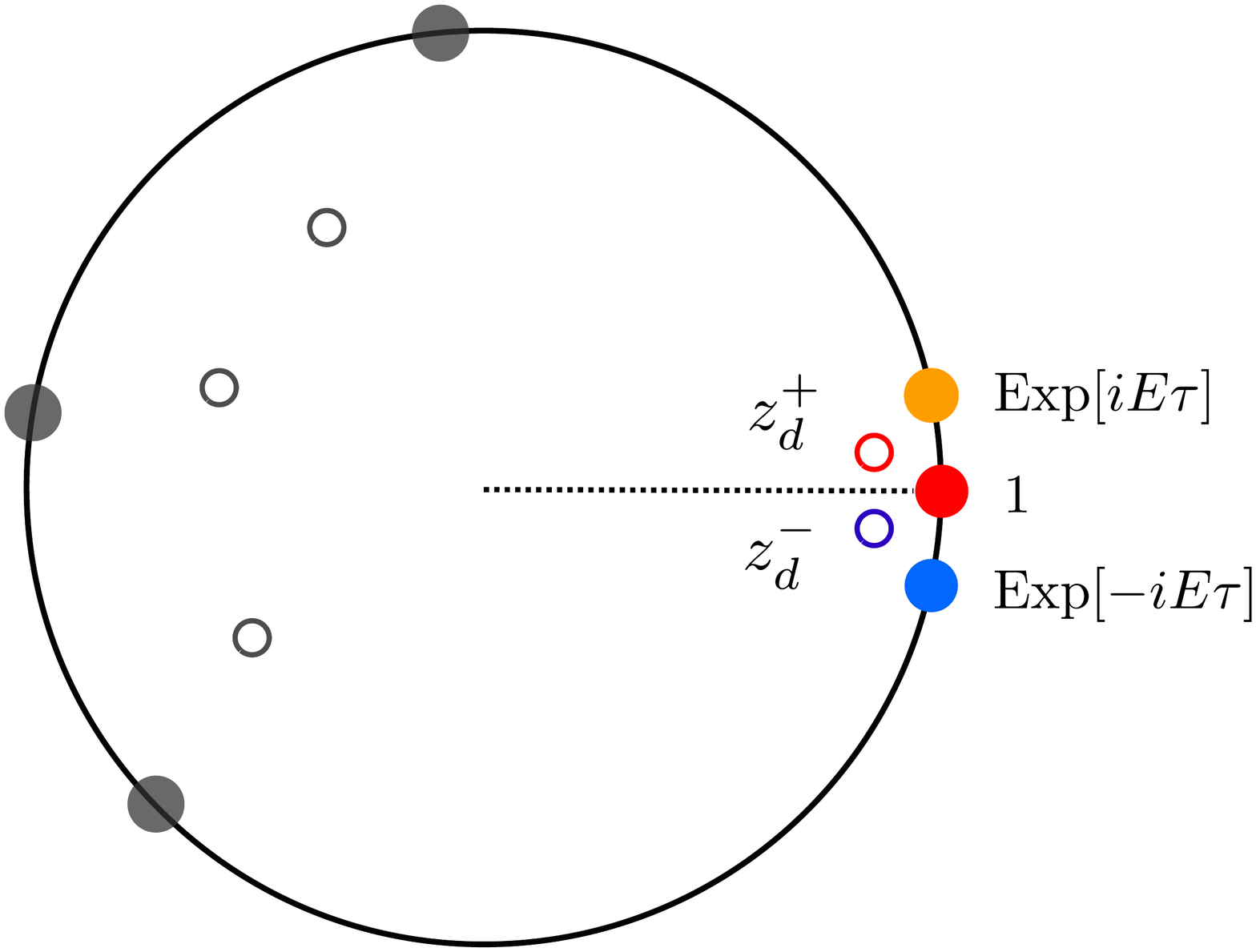}
\caption{(Color online)
Three merging  charges on the unit circle create a dimer: two zeros 
denoted $z_{d} ^{\pm}$.
}
\label{figTriple}
\end{figure}

To start with the calculation, we reduce the number of free parameters. We consider the symmetric case first, 
namely the three energy levels are assigned as $E_0, E_\pm=\pm E$. 
When the corresponding phases are merging, we rewrite $e^{iE_\pm\tau}=e^{\pm i \delta}$,
These charges imply that two zeros can be found close
to the  unit circle 
$z_d^\sigma$ with  $\sigma=\pm$, see schematics Fig. \ref{figTriple}.
Inserting the expansion 
$z_d^\sigma=1-A_\sigma\delta-B_\sigma\delta^2$ in Eq. (\ref{eqSM06})
we find to third order in $\delta$
\begin{equation}\label{AB}
\begin{aligned}
    A_{\sigma} &= -i\sigma {\sqrt{p_0\over p_0+2p}} \\
    B_\sigma &= B = {p_0+p \over 2(p_0+2p)} { -{p \over (p_0 + 2 p)^2} \sideset{}{'}\sum{p_j \over e^{iE_j\tau}-1}} 
\end{aligned}
\end{equation}
where the primed sum excludes the  three charges indices.
Hence
\begin{equation}\label{eqsm20}
\begin{aligned}
    |z_d^\sigma|^2 &= |1-A_\sigma\delta-B\delta^2|^2
    &= 1- {p \over (p_0+2p)^2}\delta^2 \le 1
\end{aligned}
\end{equation}
%
%
%

Recall the off-diagonal term ${\cal M}$ in Eq. (\ref{eq08}), substituting
$z_d^\sigma=1 - A_\sigma\delta - B\delta^2$ into it yields
\begin{equation}\label{Mnumber}
\begin{aligned}
    &\mathcal{M}\big|_{\delta\to0^+} = 2\bigg[-2 + {1 \over 1-z_d^+(z_d^-)^\ast} + {1 \over 1-(z_d^+)^\ast z_d^-}\bigg]_{\delta\to0^+} \\
    &= -4 + 2\bigg[{ 1 \over (B + B^\ast-A_+^2)\delta^2 + 2A_+\delta } + c.\,c. \bigg]_{\delta\to0^+} \\
    &=
    -3 + { B + B^\ast \over |A_+|^2} = -2 + {p \over p_0(p_0+2p)}
\end{aligned}
\end{equation}
which  is finite as mentioned. 
Therefore, we can indeed neglect ${\cal M}$ in Eq. (\ref{eq08})
as $|\tau(\bar{E}_1 - \bar{E}_2)|\to 0$. 
Then plugging Eq. (\ref{eqsm20}) into Eq. (\ref{eq08}) yields Eq. (\ref{eq09}).
~\\

\subsection{Zeno regime}\label{b4}
%
\begin{figure}
\centering
\includegraphics[width=0.48\textwidth]{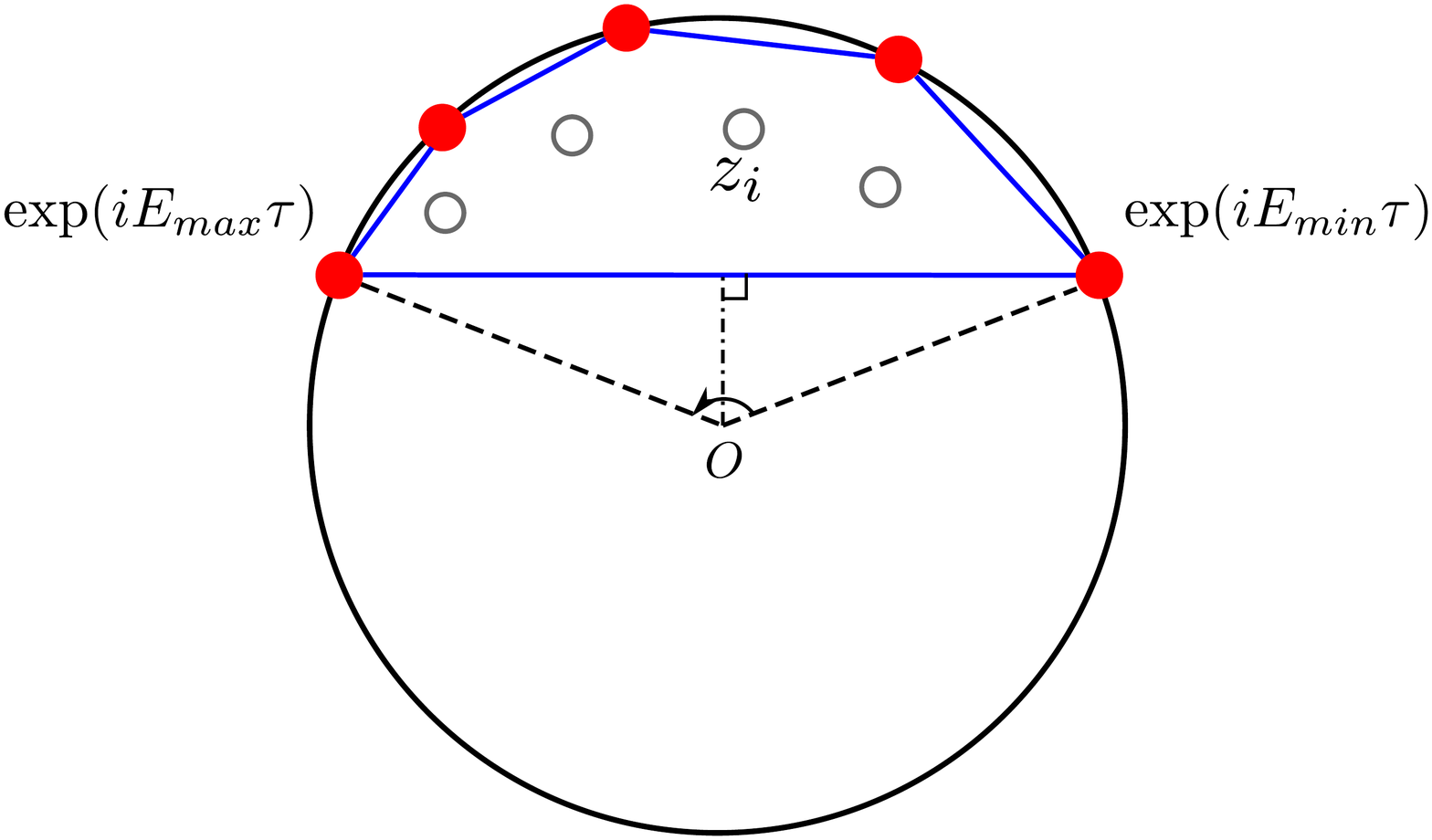}
\caption{(Color online)
In the limit of small $\tau$ the charges on the unit circle will coalesce.
As shown the zeros are within a convex hull whose vertices are the charges. 
When $\tau\to 0$ these zeros will approach the unit circle and thus the
fluctuations of $n$ are large. The calculations of the zeros, while
a possibility for small systems  with numerics, 
is non-trivial since we are dealing with a many
charge theory. Still we can obtain a useful bound for the fluctuations
using geometrical arguments. 
}
\label{figZ}
\end{figure}

We investigate the Zeno regime, where $\tau$ is small though finite,
seeking a lower bound for the fluctuations. 
We can generally rewrite the second line in Eq. (\ref{eq02}) as 
\begin{equation}
    \mbox{Var}(n) = \sum_{j,k}^{w-1} {\cal V}_{jk}
\end{equation}
where 
\begin{equation}
{\cal V}_{jk}={z_j z_k^\ast \over 1-z_j z_k^\ast} + {z_j^\ast z_k \over 1-z_j^\ast z_k}.
\end{equation}
Let $z_j=r_{j}e^{i\theta_j}$ and then
\begin{equation}\label{vjk}
{\cal V}_{jk} = -1 + {1 - r_j^2 r_k^2 \over 1 + r_j^2 r_k^2 - 2 r_j r_k \cos \Delta\theta_{jk}}
\end{equation}
where $\Delta\theta_{jk}=|\theta_{j}-\theta_{k}|$. 
As presented in Fig. \ref{figZ}, we have the bounds 
$\cos(\Delta E_\text{m} \tau/2) \le r_{j} < 1$, $0 \le \Delta\theta_{jk} < \Delta E_\text{m} \tau$ with $\Delta E_\text{m} = E_{\rm max}-E_{\rm min} $ and $E_{\rm max}/E_{\rm min}$ is the  maximum or minimum  of the discrete
energy levels.  Recall that  we consider finite systems where the spectrum is
bounded, energy levels are discrete, and hence 
$\langle n \rangle = w$ finite. 
For the diagonal terms, namely when $j=k$ every element 
\begin{equation}
    {\cal V}_{jj} \ge {2 r_j^2 \over 1-r_j^2}\bigg|_{r_j^2=\cos^2(\Delta E_{\rm m}\tau/2)} = 2\cot^2(\Delta E_\text{m} \tau/2)
\end{equation}
and for the off-diagonal terms ($j\neq k$),
\begin{equation}
    {\cal V}_{jk} > -1
\end{equation}
since the second term in the right-handed side of Eq. (\ref{vjk}) is always positive.
Thus the sum is bounded as 
\begin{equation}
    \mbox{Var}(n) = \sum^{w-1}_{j,k}{\cal V}_{jk} > 2 (w-1) \cot^2(\Delta E_\text{m} \tau/2) - (w-1)(w-2)
\end{equation}
which is Eq. (\ref{eq10}).
~\\

\section{Details on the examples}\label{appC}

\subsection{Interacting two-boson model}\label{c1}

Two bosons in a Josephson tunneling junction, governed by the Hamiltonian Eq. (\ref{bosonsH})
%
%
is our first example. 
 We focus here on the quantum return problem of $\ket{\psi_{{\rm in}}}=\ket{2,0}$,
namely we consider a detector that records two particles on the left
(other measurements are of course possible and they will be treated elsewhere).
The energy levels of this system are
\begin{equation}\label{energybosons}
  \begin{aligned}
    E_{0} &= 3U-\sqrt{U^2+J^2}, \quad
    E_1 &= 4U,\\ 
    E_2 &= 3U+\sqrt{U^2+J^2}. 
  \end{aligned}
\end{equation}
The subscript $1/2$ means first/second excited state.
Since there are three energy levels here, the average of $n$ is $\langle n\rangle=w=3$, except for special $\tau$'s, $U$'s and $J$'s, see details below.
As shown in Fig. \ref{fig3} we investigate $\mbox{Var}(n)$ 
varying the on-site interacting energy $U$ (while $\tau,J$ are fixed).
In this example we find that when $U$ becomes large
 one of the overlaps approaches zero (one charge theory)
but at the same time two charges merge (two-charge theory), so we have two effects taking place at the same time.

\begin{figure}
\centering
\includegraphics[width=0.35\textwidth]{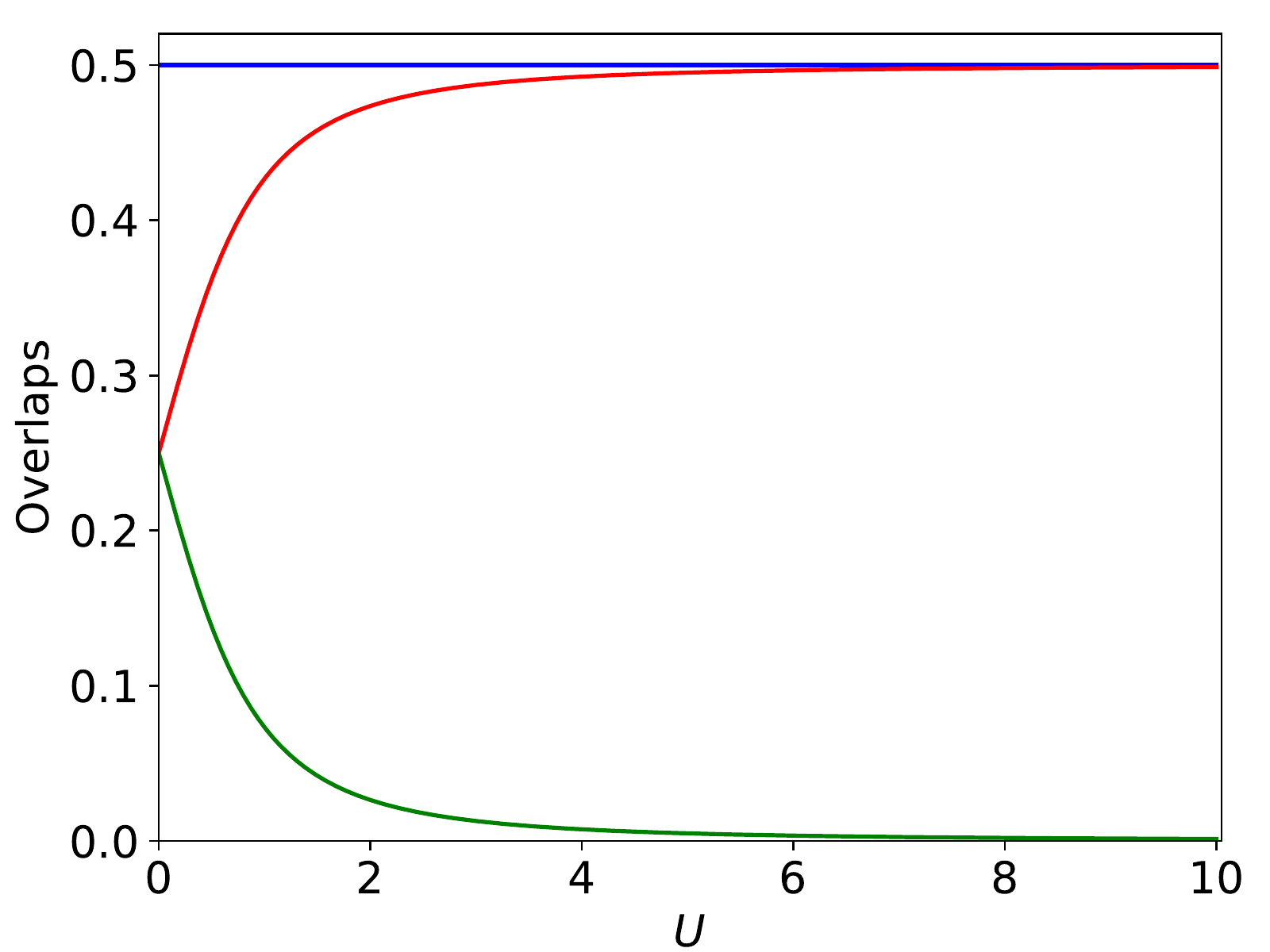}
\caption{(Color online)
The overlaps $p_j$ versus interacting strength $U$ for the interacting two-boson model with $J=1$. The green curve represents $p_{0}$, and the blue/red is $p_1/p_2$. As shown, $p_{0}$ decays along with the increase of $U$. When $U$ is large, $p_{0}$ is almost $0$, while $p_1,p_2$ are finite.
}
\label{overlaps}
\end{figure}

Specifically, $\exp(iE_1\tau)=\exp(iE_2\tau)$  when $U =J^2\tau/4\pi k - \pi k/\tau$ when $k=1,2...$ and notice that this merging may take place only 
when $J \tau>2 \pi$ (we consider only $U>0$). 
The merging of two other  phases  $\exp(iE_1\tau)=\exp(iE_{0} \tau)$ 
happens when $U= -J^2\tau/4\pi k + \pi k/\tau$. 
Finally, $\exp(iE_2\tau)=\exp(iE_{0}\tau)$ will merge when $U= \sqrt{\pi^2 k^2/\tau^2-J^2}$.
Since we choose $J=1,\tau=3$ only the last two cases
are relevant in Fig. \ref{fig3} in the main text.
Note that the overlaps of $\ket{E_j}$ with $\ket{2,0}$ are also parameterized by $U,J$:
\begin{equation}
  \begin{aligned}
    p_{0} &= {J^2 \over 4 \sqrt{U^2+J^2}(\sqrt{U^2+J^2}+U)}, \quad
    p_1 = 1/2,\\
    p_2 &= {J^2 \over 4 \sqrt{U^2+J^2}(\sqrt{U^2+J^2}-U)}. 
  \end{aligned}
\end{equation}
We plot the overlaps $p_j$ changing with $U$, as shown in Fig. \ref{overlaps}.
So as mentioned in the main text we also have one-charge dominance here. 
%
%
The critical $U$'s
give $|E_1-E_{0}|\tau=2\pi k$ or $|E_2-E_{0}|\tau=2\pi k$, and these
are described  by the single-charge theory Eq. (\ref{eq04}).
Note that when $U$ becomes very large, $E_1 \simeq E_2$, as mentioned in the main text.
This  leads to large fluctuations of $n$ coming from two sources: 
two phases are merging on the unit circle, 
together with single charge effects since $p_0$ is also small.
Thus the final approximation is the sum of Eq. (\ref{eq04}) and Eq. (\ref{eq06}), namely,
\begin{widetext}
\begin{equation}\label{sumoftwoandone}
\begin{aligned}
  \text{Var}(n) \sim& {1 + \big( \sum_{j=1,2} p_j\cot[({E}_j-{E}_{0})\tau/2] \big)^2 \over 2\,p_{0}}
  &+ 2\;{(p_1+p_2)^3 \over  p_{1}p_{2}}{1 \over \tau^2 (\bar{E_{1}}-\bar{E_{2}})^2}
\end{aligned}
\end{equation}
\end{widetext}
In Fig. \ref{fig3}
we show that the approximation perfectly matches the exact results when $U$ is large.
~\\

\subsection{The eight-site ring}\label{c2}

\newcommand{\rom}[1]{\uppercase\expandafter{\romannumeral #1\relax}}
The Hamiltonian of the eight-site ring model Eq. (\ref{eq07}),
gives the energy levels 
$E_k=2-2\cos(\pi k/4)$, i.e., $E_0=0$, $E_{{\rm \rom{1}}}=E_1=E_7=2-\sqrt{2}$,
$E_{{\rm \rom{2}}}=E_2=E_6=2$, $E_{{\rm \rom{3}}}=E_3=E_5=2+\sqrt{2}$ 
and $E_{{\rm \rom{4}}}=E_4=4$ 
when $\gamma=1$. 
So here as
mentioned,  $\langle n\rangle=w=5$ except for special sampling times
given by $\tau= \pi j/2$, $2\pi j/(2+\sqrt{2})$, $\pi j/\sqrt{2}$, $\pi j$, $\sqrt{2}\pi j$ and $2\pi j/(2-\sqrt{2})$ with $j$ an integer.
The corresponding eigenstates are
\begin{equation}
\begin{aligned}
    \ket{E_k}&=\sum_{x=0} ^{7} e^{i \pi x k/4}\ket{x}/2\sqrt{2} \\
\end{aligned}
\end{equation}
hence $|\langle \psi_{\rm in}| E_k \rangle|^2=1/8$ with $k=0,1,\cdots,7$, 
however we have degeneracy so we define: $p_0=p_{\rm \rom{4}}=1/8$, $p_{\rm \rom{1}}=1/4$,
$p_{\rm \rom{2}}=1/4$, $p_{\rm \rom{3}}=1/4$.
Thus on the unit circle we have five charges: two with charge $1/8$ and three
with charge $1/4$, these merge for the mentioned
special sampling times $\tau$.

We now explain how to get the approximations presented in Fig. \ref{fig2}(F), 
for that  we find $\text{Var}(n)$ close to resonances.
\begin{itemize}
\item[1.]
When $\tau$ is equal to $\pi$, $2 \pi$ or $\sqrt{2}\pi$  
 there are only three phases on the unit circle. Hence close to these sampling times we get the blow-up
of $\mbox{Var}(n)$. 
Specifically in these cases  $(e^0,e^{i2\tau},e^{i4\tau})$ and $(e^{i(2-\sqrt{2})\tau},e^{i2\tau},e^{i(2+\sqrt{2})\tau})$ merge.
Hence in this case we use the triple-charge theory 
Eq. (\ref{eq09}). 
These cases are the peaks colored in green in Fig. \ref{fig2}(F).

For the first case, consider $\tau\simeq\pi$, the parameters in Eq. (\ref{eq09}) are, 
$p_0=p_{2,6}=1/4$, $p=1/8$, 
so 
$\text{Var}(n) \sim 2/(\pi-\tau)^2$; 
while when $\tau\simeq2\pi$ we get $\text{Var}(n) \sim 2/(2\pi-\tau)^2$.
The configurations of charges and zeros are shown in Fig. \ref{fig2}(A-D).
For the case $\tau\simeq\sqrt{2}\pi $
with $p_0=p=1/4$, $\text{Var}(n) \sim 9/[2(\sqrt{2}\pi-\tau)^2]$.

\item[2.]
When $\tau\simeq 2\pi j/(2+\sqrt{2})$ with $j=1,2,3$, we have two zeros separately approaching to the unit circle as we have two couple of phases/charges merging
on the unit circle. These correspond to the three blue peaks in Fig. \ref{fig2}(F).
Using Eq. (\ref{eq06}) we find the approximation 
$\mbox{Var}(n)\sim 27/[4(2\pi j-(2+\sqrt{2})\tau)^2]$.
Here the contribution from the two zeros add up. 
We use $j=1,2,3$ since we consider $\tau$ in the interval $(0,2 \pi)$,
see Fig. \ref{fig2}.

\item[3.]
When $\tau \simeq \pi/2, \pi/\sqrt{2}, 3\pi/2$, we have the two-charge theory,
namely  one zero approaching the unit circle 
 Eq. (\ref{eq06}), these correspond to the 
three peaks colored in pink in Fig. \ref{fig2}(F).
We find the corresponding approximations $\mbox{Var}(n)\sim 1/[8(\pi/2-\tau)^2], \mbox{Var}(n)\sim 1/[2(\pi/\sqrt{2}-\tau)^2]$ and 
$\mbox{Var}(n)\sim 1/[8(3\pi/2-\tau)^2]$ respectively.

\item[4.]
When $\tau\simeq 0$, namely in the Zeno regime, we apply Eq. (\ref{eq10})
 which gives $\mbox{Var}(n)> 8\cot^2(2\tau) - 12$ as a lower bound.
\end{itemize}

All this information is presented Fig. \ref{fig2}(F) perfectly
matching the exact solution. 
~\\

\subsection{Symmetry breaking in a ring system with defect}\label{c3}
%
\begin{figure*}[htbp]
\centering
\includegraphics[width=0.8\linewidth]{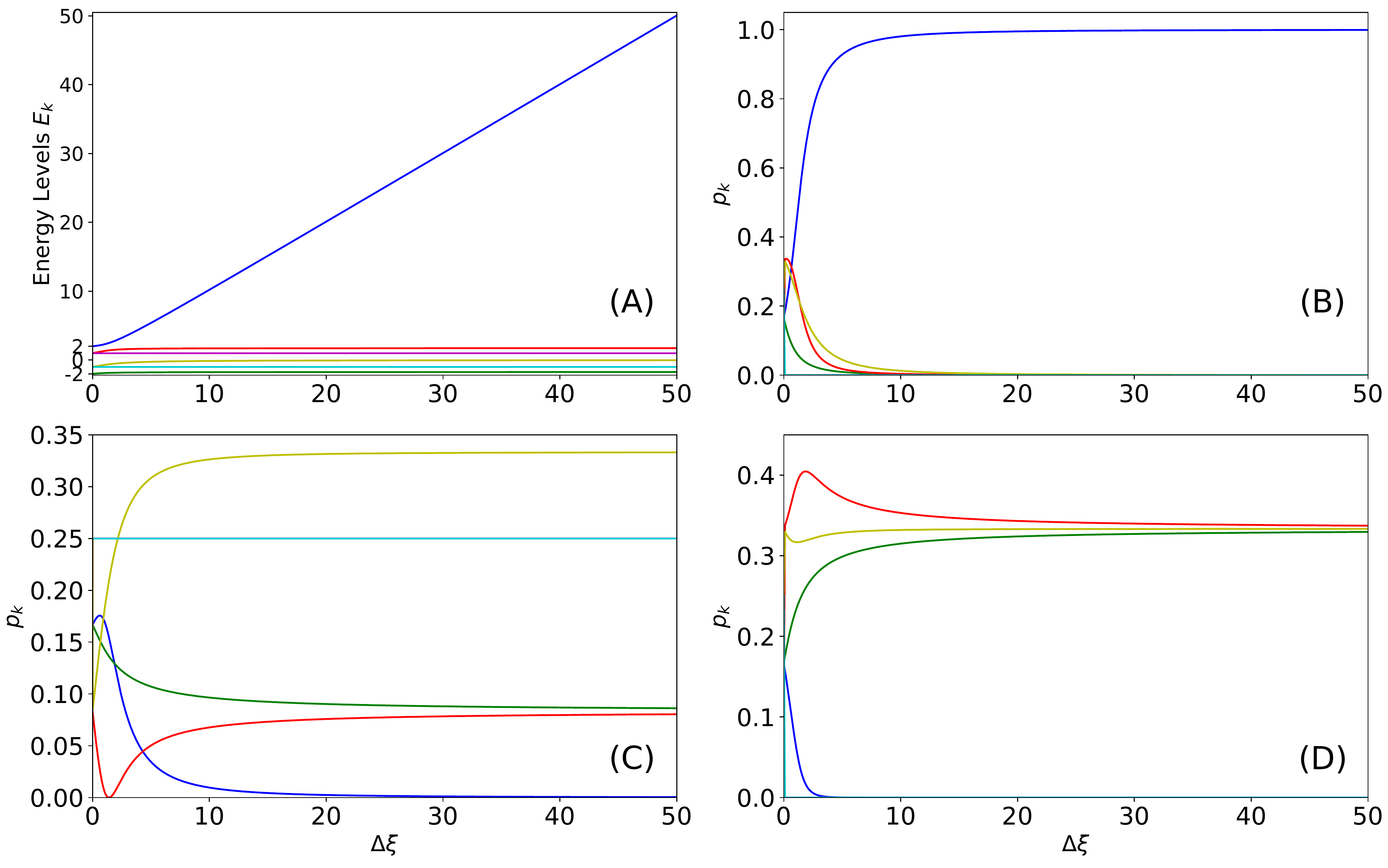}
\caption{(Color online)
(A) The energy levels, (B-D) the overlaps $p_k$ of eigenstates $\{ \ket*{E_k} \}$ and the detected state 
versus the defect strength $\Delta\xi$ for six-site ring with one defect.
See Fig. \ref{fig6s} for notations and schematics of this system.
Here $\gamma$ in Eq. (\ref{eq1}) is set to $1$, and $\Delta\xi$ ranges from $0$ to $50$.
(B) when the detector set at site $x=0$, 
where the defect is located;
(C) when the detected/initial state is $\ket*{1}$, the neighbor of the defect;
(D) when the state $\ket*{3}$ is the target.
Notice the color coding, e.g. color blue for an energy level and the same color for the corresponding overlap.
See further details in the text.
}
\label{fig6o}
\end{figure*}

The final example in the main text is governed by the Hamiltonian Eq. (\ref{eq1}).
The defect will increase the original energy levels of the clean system.
As mentioned, the defect removes the degeneracy of the system. 
For $\Delta\xi = 0$ the number of distinct energy levels is four 
while when $\Delta\xi \neq 0$ we have six levels. 
Fig. \ref{fig6o}(A) shows the influences of the defect on the energy levels.

In the following we consider the mean of $n$ and its fluctuations for three different cases. 
First when the initial state is $\ket*{\psi_\text{in}} = \ket*{0}$ (namely on the defect) 
and then when $\ket*{\psi_\text{in}} = \ket*{1}$ (the nearest neighbor),
$\ket*{\psi_\text{in}} = \ket*{3}$ (the opposite), see Fig. \ref{fig6s} for notations. 
Figs. \ref{fig6o}(B-C) describe the corresponding overlaps $p_{k}$ 
as a function of the defect $\Delta\xi$, 
and Fig. \ref{fig6d} gives the fluctuations of $n$ versus $\Delta\xi$, 
from which one find gigantic variance around some special $\Delta\xi$.
Due to symmetry breaking the return time is not translation invariant.

As mentioned, with the introduction of the defect at site $x=0$, 
the degenerate energy levels of the original clean system split.
For the ``perfect'' six-site ring, i.e. $\Delta\xi = 0$,
$E(k) = -2\gamma \cos(\pi k/3)$ with $k=0,1,2,3,4,5$,
so there are four energy levels $-2\gamma(1), -\gamma(2), \gamma(2), 2\gamma(1)$, 
and the numbers in parentheses are degeneracies.
With the defect, we have six non-degenerate energy levels.
To determine the winding number $w$ we need to find the number of distinct energy levels 
whose overlaps with the detected state are not zero. 
We notice that two exact solutions to $|H-\lambda I|=0$ of the Hamiltonian Eq. (\ref{eq1}) are:
$\gamma, -\gamma$, 
and the corresponding eigenvectors are:
$\{0, 1/2, -1/2, 0, 1/2, -1/2 \}$, $\{0, -1/2, -1/2, 0, 1/2, 1/2\}$,
which indicates one property of the model: reflection symmetry.
Therefore, 
when the detected state is $\ket*{0}$ or $\ket*{3}$, 
owing to the vanished charges/overlaps $p_{(\gamma)}, p_{(-\gamma)}$,
the mean of $n$ or the winding number is $4$, the same as the perfect ring.
This is valid for nearly any $\Delta \xi$ the exceptions are singular points 
where two phases of the problem merge.
Specifically, for the former case $\ket*{\psi_\text{in}} = \ket*{0}$,
there are three weak charges, and one strong charge corresponding to the largest energy, 
as $\Delta\xi$ becomes large,
see Fig. \ref{fig6o}(B).
Then there are three zeros close to the three weak charges, 
making dominant contributions to the variance.
The final approximation is the summation of all three $2|z_i|^2/(1-|z_i|^2)$, or Eq. (\ref{eq04}):
\begin{equation}\label{asympVa}
	\text{Var}(n)	\sim
						\sum_{k=1,3,5} {1 \over 2 p_k}\bigg\{ 1 + \bigg[ p_6 \cot[(E_6 - E_k)\tau/2] \bigg]^2 \bigg\}
\end{equation}
Here the indices are from sorting energy levels from small to larger, 
i.e. $E_1$ is the ground energy, $p_1$ is the overlap of its eigenstate with $\ket*{0}$.
Inside the bracket, we only use the charge $p_6$, the strong charge, 
since other $p_k$ ($k=1,3,5$, $p_2=p_4=0$) are small 
and not merging with one another under our case,
making non-significant contributions. 
For the latter case, where the target is $\ket*{3}$, 
we have simply one weak charge, corresponding to the largest energy, 
when $\Delta\xi$ becomes large.
It is easy to understand, since its eigenvector has a large overlap with $\ket*{0}$, 
the other overlaps with other position states should be effectively small.
As a result, applying Eq. (\ref{eq04}) gives a perfect asymptotic description for the fluctuations of $n$ in Fig. \ref{fig6d}(C).
Note that all the peaks in the variance figures for the two cases are caused by the merging of weak charge(s) and strong charge(s), when the winding number is reduced by $1$.

When the initial state is $\ket{1}$, we have a different behavior.
And this is because the location of the detector breaks the reflection symmetry of the system. 
we have six charges on the unit circle as long as $\Delta\xi > 0$.
It means that when $\Delta\xi$ changes from zero to be positive, 
there is a jump of the winding number from $4$ to $6$, 
around which we find enormous variance of $n$ as expected. 
Here we apply the two-charge theory 
to two pairs of charges' merging at different locations, 
which gives us an excellent agreement between exact solution and our theory [see Fig. \ref{fig6d}(B)].
Furthermore, from Fig. \ref{fig6o}(C), 
we see that depending on the value of $\Delta\xi$ 
we have different small charges 
(red line when $\Delta\xi \simeq \sqrt{2}$ and blue when $\Delta\xi \gg 1$). 
Notice that the charge represented by the red line decreases to $0$ when $\Delta\xi = \sqrt{2}$, 
leading to $w=5$ at this critical value.
Thus we sum over the contributions of both dominating zeros to provide an approximation.
The global asymptotic formula is
\begin{widetext}
\begin{equation}\label{GloAppr}
\begin{aligned}
	\text{Var}(n) 	\sim&
						{1\over 2p_5} 
						\bigg\{ 1+ \bigg[ \sum_{k=1,k\neq 5}^{6} p_k \cot[(E_k-E_5)\tau/2] \bigg]^2 \bigg\}
						+
						{1\over 2p_6} 
						\bigg\{ 1+ \bigg[ \sum_{k=1}^{5} p_k \cot[(E_k-E_6)\tau/2] \bigg]^2 \bigg\} \\
						&+ 2{(p_2 + p_3)^2 \over p_2 p_3} {1 \over \tau^2 (\bar{E}_2 - \bar{E}_3)^2}
						 + 2{(p_4 + p_5)^2 \over p_4 p_5} {1 \over \tau^2 (\bar{E}_4 - \bar{E}_5)^2}.
\end{aligned}
\end{equation}
\end{widetext}
%
See the matching in Fig. \ref{fig6d}(B).
Note that the clusters of five peaks in the variance of $n$ 
are due to the merging of the weak charge of the largest energy and other five charges.

\end{document}